\documentclass[9pt,twocolumn,twoside]{revtex4-1}
\usepackage{amsfonts}
\usepackage{amssymb}
\usepackage{amsmath}
\usepackage{upgreek}
\usepackage{bm}
\usepackage{physics}
\usepackage{graphicx}
\usepackage{gensymb,color}
\usepackage{xr}
\graphicspath{{images/}}

\def\eqS{\begin{equation}}
\def\eqE{\end{equation}}

\makeatletter
\def\@fnsymbol#1{\ensuremath{\ifcase#1 \or \dagger\or *\or \mathsection\or \mathparagraph\or \|\or **\or \dagger\dagger \or \ddagger\ddagger \else\@ctrerr\fi}}
\makeatother

\begin{document}

\title{Phase transitions in rolling of irregular cylinders and spheres}

\author{Daoyuan Qian}
\affiliation{Paulson School of Engineering and Applied Sciences, Harvard University, Cambridge, MA 02138, U.S.A.}
\affiliation{Centre for Misfolding Diseases, Yusuf Hamied Department of Chemistry, University of Cambridge, Lensfield Road, Cambridge CB2 1EW, U.K.}

\author{Yeonsu Jung}
\affiliation{Paulson School of Engineering and Applied Sciences, Harvard University, Cambridge, MA 02138, U.S.A.}

\author{L. Mahadevan}
\thanks{lmahadev@g.harvard.edu}
\affiliation{Paulson School of Engineering and Applied Sciences, Harvard University, Cambridge, MA 02138, U.S.A.}
\affiliation{Department of Physics, Harvard University, Cambridge, MA 02138, U.S.A.}
\affiliation{Department of Organismic and Evolutionary Biology, Harvard University, Cambridge, MA 02138, U.S.A.}





\begin{abstract}
When placed on an inclined plane, a perfect 2D disk or 3D sphere simply rolls down in a straight line under gravity. But how is the rolling affected if these shapes are irregular or random? Treating the terminal rolling speed as an order parameter, we show that phase transitions arise as a function of the dimension of the state space and inertia. We calculate the scaling exponents and the macroscopic lag time associated with the presence of first and second order transitions, and describe the regimes of co-existence of stable states and the accompanying hysteresis. Experiments with rolling cylinders corroborate our theoretical results on the scaling of the lag time. Experiments with spheres reveal closed orbits and their period-doubling in the overdamped and inertial limits respectively, providing visible manifestations of the hairy ball theorem and the doubly-connected nature of $\bm{SO(3)}$, the space of 3-dimensional rotations. Going beyond simple curiosity, our study might be relevant in a number of natural and artificial systems that involve the rolling of irregular objects, in systems ranging from nanoscale cellular transport to robotics. 
\end{abstract}

\maketitle

Rolling of regular rigid objects forms the basis of a diverse range of research directions including the rolling-bouncing transition of ellipses \cite{Kajiyama2021}, collision dynamics of rolling polygons \cite{Beunder2003}, passive `walking' robots \cite{Coleman1997}, and more recently a framework for designing 3D shapes that roll along pre-determined trajectories \cite{Sobolev2023}. Furthermore, optimal control of a rolling ball by an internally moving point mass \cite{Putkaradze2018,Putkaradze2020,Ilin2017,Putkaradze2021} is also an active field of research. While the deterministic case has been well studied for more than a century, recent studies have started to address the case of stochastic driving in such situations, as when the rolling plane is subject to random fluctuations  \cite{Goohpattader2011}. Complementing the case of the stochastic driving of a regular rolling object, one might ask about the effect of shape randomness of the body on its rolling behavior. For example, how would a slightly irregular object that has random undulations on its surface roll?

\begin{figure*}
\includegraphics[width=16cm]{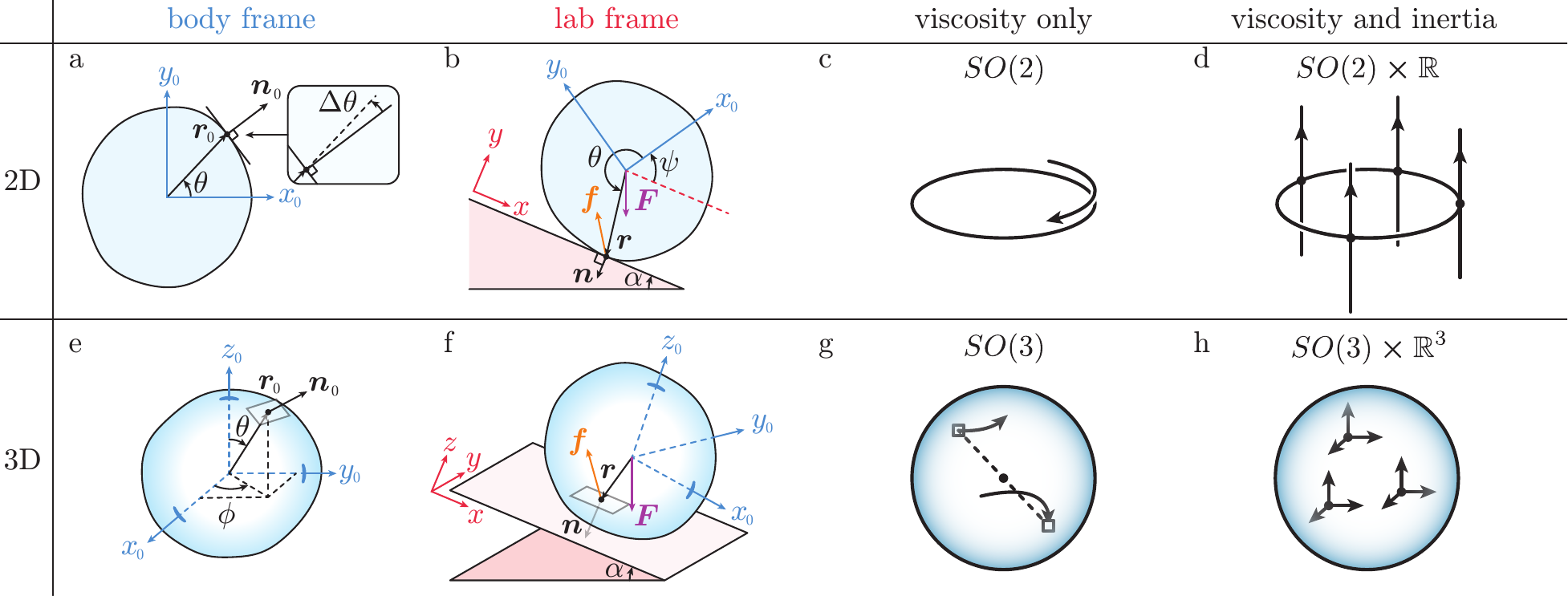}
\caption{\textbf{Setup, notations, and mathematical phase spaces of rolling motion.} \textbf{a} In 2D rolling, the contact point in the body frame is parameterised by an angle $\theta$ which serves as the main dynamic variable. The normal vector $\bm{n}_{0}$ is in general not parallel to the contact vector $\bm{r}_{0}$ and the offset is defined as $\Delta\theta$ (inset). \textbf{b} In the body frame, a body-to-lab frame rotation angle $\psi$ specifies the 2D body's orientation. These are related to the constraint of $\bm{n}$ being perpendicular to the rolling plane. \textbf{c} The state space of 2D viscous rolling is $SO(2)$, graphically it is a 1D closed loop. \textbf{d} For 2D inertial rolling, an extra real number is needed to encode the angular speed. \textbf{e} The body frame of 3D rolling requires two angles $\theta$ and $\phi$ to specify the contact point. $\textbf{f}$ To transform to the lab frame, a 3D rotation needs to be specified. The same constraint of $\bm{n}$ being perpendicular to the rolling plane applies. \textbf{g} We use elements of $SO(3)$ to describe 3D rotations. Each element can be represented as a 3D vector within a unit 3-ball, with antipodal points identified because they describe the same rotation. This means a trajectory hitting the bounding 2-sphere appears on the other side of the ball, as illustrated. \textbf{h} When inertia is considered in 3D rolling, 3 real numbers are needed to specify the angular velocity at each orientation.}
\label{fig:setup}
\end{figure*}

Here, we consider the rolling of irregular cylinders and spheres on a plane inclined at angle $\alpha$, and use a combination of theory, computation and experiment to show that the resulting behavior is extremely rich.  The object shape can be characterised using an angle-dependent radius $r(\theta)$ in 2D, or $r(\theta,\phi)$ in 3D, where the angles $\theta$ and $\phi$ encode which point on the object's surface, in the body-fixed reference frame, is in contact with the plane (Figure \ref{fig:setup}, a and e). Shape randomness is then characterized by allowing $r(\theta)$ and $r(\theta,\phi)$ to fluctuate around a mean value $\expval{r}$. Components of the contact vector connecting the centre of mass to the contact point in the body and lab frames are denoted $\bm{r}_{0}$ and $\bm{r}$ respectively, related by a body-to-lab frame rotation (Figure \ref{fig:setup}, b and f). At the contact point in the lab frame, the vector normal to the object's surface $\bm{n}$ is constrained to be perpendicular to the inclined plane. An external force $\bm{F}$ pointing vertically downwards is applied at the centre of mass of the object, generating a torque equal to $\bm{r}\cross\bm{F}$. Since $\bm{r}$ and $\bm{n}$ are in general not parallel, the torque varies with $\bm{r}$, and it is the central effect of shape randomness on rolling. The orientation-dependent torque due to the external force leads to all manner of interesting effects in both two and three dimensions.

Our investigations focus on the role of three key ingredients: an isotropic rotational viscous drag, the vertical external force, and isotropic inertia (rotational and linear). In the viscous regime where only the first two are relevant, the state spaces of motion correspond to dynamics on the two and three-dimensional rotation groups $SO(2)$ and $SO(3)$ (Figure \ref{fig:setup}, c and g), with each state point corresponding to a body-to-lab frame rotation, thus describing the body's orientation. No extra information is needed to characterise a state because the angular velocity is given directly as a function of orientation. In the regime where inertia is significant, angular acceleration becomes relevant so the angular velocities have to be stitched onto the orientation space as extra fibres to represent the full state, expanding $SO(2)$ and $SO(3)$ to $SO(2)\cross\mathbb{R}$ and $SO(3)\cross\mathbb{R}^{3}$ respectively (Figure \ref{fig:setup}, d and h).

Describing the rolling motion in these state spaces directly reveals the interplay between the state space dimension and possible dynamics. In 2D viscous rolling, the space $SO(2)$ is a closed 1D loop (Figure \ref{fig:setup} c) so a rolling body visits every possible state unless it halts in one of the states. This precludes the co-existence of a halting state and a cruising trajectory (where the object rolls indefinitely), and a second order halting-cruising transition is expected when the ramp angle $\alpha$ changes. In 2D inertia rolling, the state space $SO(2)\cross\mathbb{R}$ is 2-dimensional (Figure \ref{fig:setup} d) so halting states and cruising trajectories can co-exist, leading to first order phase transitions instead. Similarly, first order phase transitions are expected for 3D viscous and inertial rolling, because the state spaces are 3 and 6-dimensional respectively (Figure \ref{fig:setup} g, h). We explicitly demonstrate these in 2D and 3D rolling. For each scenario, we derive equations of motion through force and torque balances, analyse phase-transition-related phenomena through analytical theory and numerical simulations, and finally verify our predictions in experiments. Additionally, periodic rolling is uncovered in 3D viscous rolling, which undergoes period-doubling when inertia is considered. These behaviours can be explained using the hairy ball theorem and the double-connectedness of $SO(3)$.

\begin{figure*}[htb]
\includegraphics[width=16cm]{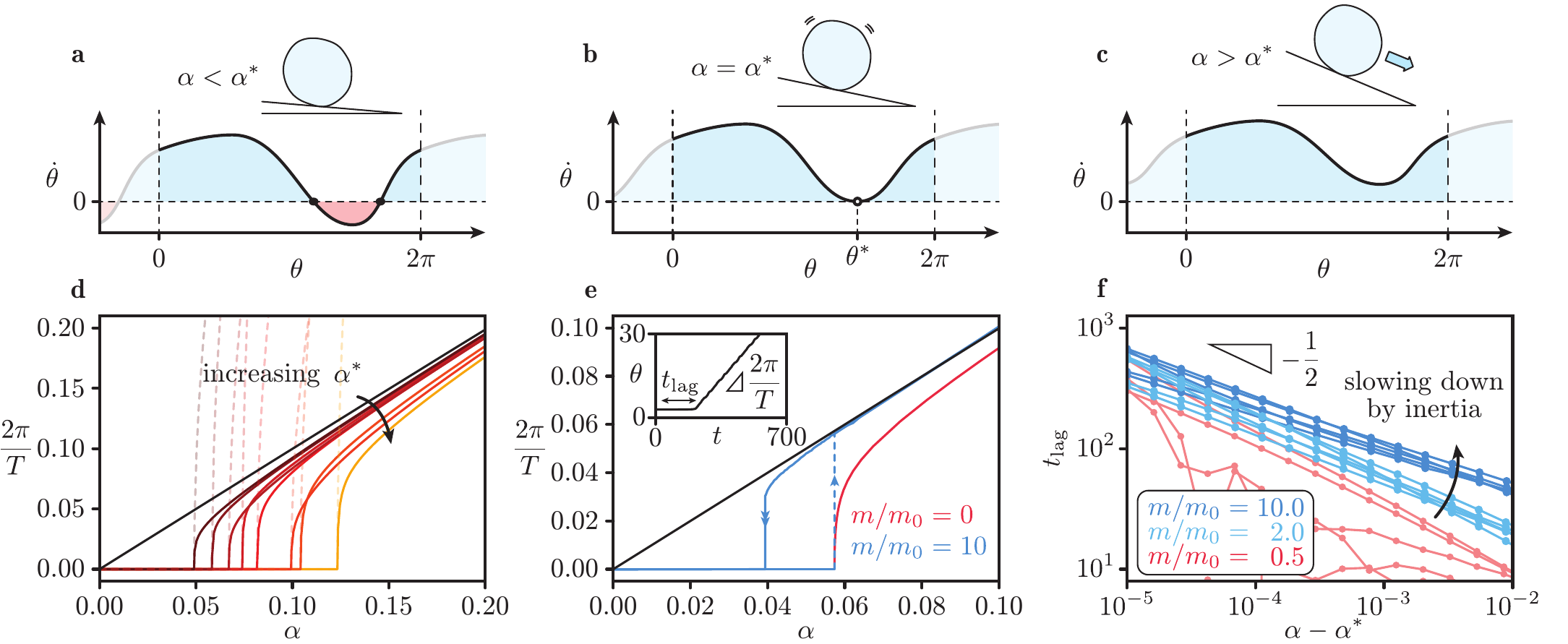}
\caption{\textbf{Critical behaviour in 2D rolling.} \textbf{a}, \textbf{b} and \textbf{c} are illustrations showing the instantaneous contact angular velocity $\dot\theta$ at different contact points $\theta$ in the viscous limit. \textbf{a} For a specific instantiation of the disk, when the angle of inclination $\alpha$ is smaller than the critical angle $\alpha^{*}$ there exists regions where $\dot\theta<0$, leading to eventual halting in the viscous limit. \textbf{b} Criticality occurs when $\alpha=\alpha^{*}$ such that $\dot\theta=0$ at a single point. \textbf{c} As $\alpha$ increases beyond $\alpha^{*}$ we have $\dot\theta>0$ so the disk rolls on indefinitely. Plots of $\dot{\theta}$ in \textbf{a} to \textbf{c} are illustrations only. \textbf{d} Simulation results for 2D viscous rolling. Coloured solid lines: plot of $2\pi/T$ as $\alpha$ is increased, for a collection of different random disks. The colours indicate the magnitude of $\alpha^{*}$ with lighter colours corresponding to larger $\alpha^{*}$. Dashed lines are plots of the analytical square-root scaling forms which are quenched by the rest of the integrand. The black solid line represents the angular velocity of a perfect disk. \textbf{e} With inertia, the transition becomes first-order (blue lines), and increasing $\alpha$ beyond $\alpha^{*}$ leads to a discontinuous jump (dashed blue line with single arrow). Hysteresis effect can be observed by simulating the rolling motion with a gradually decreasing $\alpha$, and a non-zero average angular speed is observed for $\alpha<\alpha^{*}$, before returning to 0 through a continuous, but extremely sharp, transition (double arrow). Close to criticality, the system is characterised by a lag time $t_{\text{lag}}$ where the contact angle barely moves for a very long time (inset). \textbf{f} The scaling of $t_{\text{lag}}$ with $\alpha-\alpha^{*}$ is identical to that of $T$ in the viscous limit, since they are effectively governed by the same dynamic equations.}
\label{fig:2D_Comp}
\end{figure*}

\section*{2D rolling}
\subsection*{Equations of motion}

In 2D rolling, the time-evolution of two main variables are of interest: the contact point angle $\theta(t)$ measured in the body frame (Figure \ref{fig:setup} a) and the body-to-lab rotation angle $\psi(t)$ (Figure \ref{fig:setup} b). We use dots to represent time derivatives $\dot{\theta}(t)\equiv\frac{\dd}{\dd t}\theta(t)$, $\dot{\psi}(t)\equiv\frac{\dd}{\dd t}\psi(t)$, and so on. Torque and force balances in the lab frame around the centre of mass, as well as the non-slip constraint, give \cite{Levi2014}
\eqS
\begin{split}
I\ddot{\psi}&=(\bm{r}\cross\bm{f})\cdot\bm{\hat{k}}-\gamma\dot{\psi},\\
m\dot{\bm{v}}&=\bm{f}+\bm{F},\\
\bm{0}&=\bm{v}+\bm{r}\cross(-\dot{\psi}\bm{\hat{k}}).
\end{split}
\label{eq:balance2D}
\eqE  
Here $I=\frac{1}{2}m\expval{r}^{2}$ is the moment of inertia of the disk with mass $m$, $\bm{r}$ is the vector connecting the body centre to the contact point, written as $\bm{r}=r(\theta)\cdot[\cos(\psi+\theta),\sin(\psi+\theta),0]^{T}$ (Figure \ref{fig:setup} b) where the $z$-component is written out explicitly. $\bm{f}$ is the contact force acting on the disk due to the plane, $\bm{\hat{k}}$ is the unit vector pointing out of the 2D plane, $\gamma$ is the rotational friction, $\bm{v}$ is the velocity of the centre of mass in the lab frame, and $\bm{F}$ is the gravitational force given by $\bm{F}=F\cdot[\sin\alpha,\cos\alpha,0]^{T}$ with $F$ its magnitude.

Equations (1) specify how $\psi(t)$ evolves, but we also need to propagate $\theta(t)$ in time, so a relation between $\theta(t)$ and $\psi(t)$ is needed. This can be done using the condition that the normal vector $\bm{n}$ is always perpendicular to the rolling plane (Figure \ref{fig:setup} b). If $\Delta\theta$ is the angular difference between the contact vector and normal vector in the body frame (Figure \ref{fig:setup} a, inset), given by $\tan\Delta\theta=\frac{r'(\theta)}{r(\theta)}$ (SI section 1A, Figure S1), where $r'(\theta)\equiv\frac{\dd}{\dd\theta}r(\theta)$, we are lead to the relation $\psi(t)+\theta(t)=\frac{3}{2}\pi+\Delta\theta(t)$ at all times.  Using these relations we derive the map between $\psi$ and $\theta$ (SI section 1A)
\eqS
\psi=-\int\kappa(\theta)\sqrt{r^{2}(\theta)+r(\theta)r'(\theta)}\dd\theta
\eqE
where $\kappa(\theta)=\frac{r^{2}(\theta)+2r^{\prime2}(\theta)-r(\theta)r''(\theta)}{\sqrt{r^{2}(\theta)+r^{\prime2}(\theta)}^{3}}$ is the curvature of the disk at material coordinate $\theta(t)$. Physically this means when the disk rolls, the distance rolled expressed using either $\theta$ or $\psi$ should be equal (with the extra minus sign coming from the definition of $\psi$). Using the $\theta-\psi$ relations we can re-formulate the equations of motion in terms of $\theta$ alone [since $r=r(\theta)$], and a single equation governing $\theta$ can be derived (SI section 1A):
\eqS
\begin{split}
\ddot{\theta}=&\frac{1}{I+mr^{2}}\left[
\frac{Fr (r\sin\alpha-r'\cos\alpha)}{\kappa \left(r^2+r^{\prime2}\right)}-\gamma  \dot\theta \right]\\
&-\left[\frac{I+mr^{\prime2}}{I+mr^{2}}r(r\kappa)'+\kappa r'r''+\frac{mr^{2}}{I+mr^{2}}\left(r^{2}\kappa\right)'\right.\\
&\quad\quad\left.+\frac{I}{I+mr^{2}}r^{\prime2}\kappa'\right]\frac{\dot{\theta}^{2}}{\kappa(r^{2}+r^{\prime2})}.
\end{split}
\eqE
The driving force $\bm{F}$ enters the dynamics via the term proportional to $(r\sin\alpha-r'\cos\alpha)$, and we show in the following section that singular behaviours arise when it vanishes for some $\theta$.
\\

\subsection*{The transition to rolling in the overdamped regime }

\begin{figure*}
\includegraphics[width=16cm]{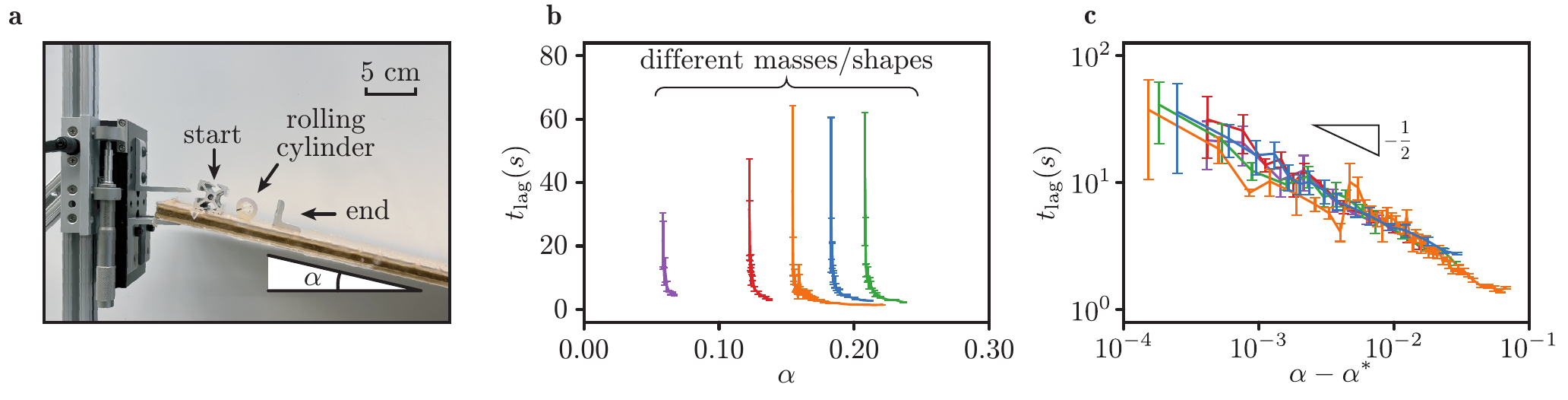}
\caption{\textbf{Validation of the $\bm{-\frac{1}{2}}$ scaling of $\bm{t}_{\textbf{lag}}$.} \textbf{a} Experimental setup. The inclination angle $\alpha$ is slowly decreased using a micrometer screw-gauge, and the time taken for the cylinder to roll over a small distance is recorded as $t_{\text{lag}}$. \textbf{b} Plot of $t_{\text{lag}}$ as a function of $\alpha$, where each run is assign a different colour. The mass or the shape of the cylinder varies across runs, leading to the singularity appearing at different $\alpha$'s. $\alpha^{*}$ values are fitted for each run. \textbf{c} Log-log plot of $t_{\text{lag}}$ against $\alpha-\alpha^{*}$ where an $\alpha^{*}$ is fitted for each run. The $-\frac{1}{2}$ scaling is observed.}
\label{fig:2D_Expt}
\end{figure*}

To separate viscous (overdamped) and inertial regimes, we note that the constants $F$, $\gamma$, and $\expval{r}$ jointly set the time and mass scales
\eqS
\begin{split}
t_{0}\equiv\frac{\gamma}{F\expval{r}},\quad m_{0}\equiv\frac{\gamma^{2}}{F\expval{r}^{3}}.\\
\end{split}
\eqE
Expressing the equation of motion in dimensionless units reveals that the first square bracket of Equation (3) has a pre-factor proportional to $\frac{m_{0}}{m}$ (SI section 1A). The viscous limit is achieved by setting $m\ll m_{0}$, leading to
\eqS
\dot\theta=\frac{Fr}{\kappa(r^{2}+r^{\prime2})}(r\sin\alpha-r'\cos\alpha).
\eqE
Plotting $\dot{\theta}$ for a fixed realisation $r(\theta)$ at various $\alpha$ reveals three distinct scenarios within one period: $\dot{\theta}=0$ for 2 or more values of $\theta$ (Figure \ref{fig:2D_Comp} a), $\dot{\theta}=0$ at a single $\theta$ (Figure \ref{fig:2D_Comp} b), and $\dot{\theta}\neq0$ (Figure \ref{fig:2D_Comp} c). These zeros in $\dot\theta$ translate to  (critical) transitions as a function of the average angular velocity $2\pi/T$ where the period $T$ is computed as 
\eqS
T=\int_{0}^{2\pi}\frac{1}{\dot\theta}\dd\theta.
\eqE
We see that when $\dot\theta=0$ for a single $\theta$, corresponding to a critical point, there is a transition from a stationary to a rolling state (Figure \ref{fig:2D_Comp} b). Zeros of $\dot\theta$ are given by the equation $\tan\alpha=r'/r$, and the critical angle $\alpha^{*}$, at which there is a single zero in $\dot{\theta}$ is given by (SI section 1B)
\eqS
\alpha^{*}=\tan^{-1}\left[\max_{\theta}\frac{\dd}{\dd\theta}\ln r(\theta)\right].
\eqE
Denoting $\theta^{*}$ as the contact angle where the disk halts at $\alpha^{*}$ (Figure \ref{fig:2D_Comp}b), 
and expanding $\dot\theta$ around the critical point by writing $\alpha=\alpha^{*}+\delta\alpha$ and $\theta=\theta^{*}+\delta\theta$ we can perform the integration in [6] to yield an explicit expression for the inverse period as (SI section 1B)
\eqS
\frac{2\pi}{T}\propto\begin{cases}0\quad&\text{if}\quad\alpha\leq\alpha^{*},\\
\sqrt{\delta\alpha}\quad&\text{if}\quad\alpha>\alpha^{*}.\end{cases}
\eqE
We see that the average angular velocity exhibits a divergent behavior as $\delta \alpha \rightarrow 0$, consistent with a simple supercritical bifurcation (second order phase transition). This is confirmed in numerical simulations (Figure \ref{fig:2D_Comp} d) carried out using multiple realizations of random 2D disks by writing $r(\theta)$ as a sum of Fourier modes, with dimensionless units $F=1$, $\gamma=1$ and $\expval{r}=1$, and then calculating $\frac{2\pi}{T}$ as a function of $\alpha$ (see SI section 1C for simulation details).

It might be instructive to compare the scaling of $\frac{2\pi}{T}$ for small $\delta\alpha$ to the case of static friction of an overdamped block resting on a ramp, with the frictional force proportional to the normal force. In the neighborhood of the critical angle $\alpha^{*}$, corresponding to the angle of repose, we expect the terminal velocity to scale linearly with $\delta\alpha$ instead, in contrast with the case of rolling studied here.

\subsection*{The transition to rolling in the inertial regime}

The second order transition from the halting state to the moving state when inertia is absent stems from the fact that the governing first order dynamical system has no memory. However, when inertia is present, once the disk has gone past $\theta^{*}$ for the first time, it becomes easier to cross it again and the long-time behaviour of $T$ depends on the balance of torque, friction and inertia for a range contact angles in one revolution. The average terminal angular velocity would thus assume a finite value as soon as $\alpha$ is increased beyond $\alpha^{*}$ so one expects a first-order transition (sub-critical bifurcation). We first demonstrate this in simulations by setting $m/m_{0}=10$ while fixing the shape and setting $\theta(0)=\theta^{*}$ as the initial contact point. The angle $\alpha$ is gradually increased and then decreased to 0 (Simulation details in SI section 1C). We see that there is indeed a discontinuous jump in $2\pi/T$ at the transition point $\alpha=\alpha^{*}$ (Figure \ref{fig:2D_Comp} e, single arrow) when $\alpha>\alpha^{*}$. Continuing the simulation by decreasing $\alpha$ below $\alpha^{*}$ reveals hysteresis: the rolling speed $\frac{2\pi}{T}$ does not  vanish when $\alpha=\alpha^{*}$, and in fact remains finite for a range of $\alpha<\alpha^{*}$ as the disk is now entrapped in a cruising state. This cruising state persists for smaller value of $\alpha$ (Figure \ref{fig:2D_Comp} e, double arrow),  until a second critical value of $\alpha$ is reached, wherein the solution reverts to a halting state abruptly (Figure S2). This hysteretic transition is signature of an underlying saddle-node bifurcation, and usually leads to a logarithmic divergence in the frequency instead of a power-law, i.e. $\frac{2\pi}{T}\propto-\frac{1}{\ln\delta\alpha}$ for $\delta\alpha$ measured near the new transition angle (SI section 1D).

This behavior is typical of many dynamical systems, e.g. the time for a rigid pendulum to leave the unstable, upright position \cite{Hall2013}, the transient dynamics associated with colliding wave fronts close to a nucleation solution in excitable media \cite{Argentina1997}, etc. Additionally, inspecting the $\theta(t)$ trace for $\alpha$ just above $\alpha^{*}$ reveals a macroscopic lag phase $t_{\text{lag}}$ before the onset of motion (Figure \ref{fig:2D_Comp} e, inset). This behavior is also observed in a diverse systems that exhibit first-order phase transitions, such as the failure of interdependent networks \cite{Gross2023} and protein aggregation \cite{Arosio2015}. In our system, a natural question is how $t_{\text{lag}}$ relates to physical quantities. Since $t_{\text{lag}}$ is much larger than the equilibration time to settle into the cruising state corresponding to rolling motion, we approximately have $\ddot\theta=0$, and we can approximate $\dot\theta$ using [5] so that $t_{\text{lag}}$ scales the same way as the period $T$ in the viscous limit:
\eqS
t_{\text{lag}}\propto\frac{1}{\sqrt{\delta\alpha}}.
\eqE
This scaling is observed in numerical simulations (Figure \ref{fig:2D_Comp} f). For small $m/m_{0}=0.5$, some instantiations of the disk can take a long time to reach the equilibrium state and the estimation of $t_{\text{lag}}$ becomes difficult as it is subject to strong noise; at intermediate $m/m_{0}=2$ the inverse square root scaling is captured; at higher $m/m_{0}=10$ deviations are seen again especially at larger $\alpha$, and this is because the effect of memory is stronger with larger $m/m_{0}$, which introduces additional terms in the expression of $\dot\theta$ that slows down the rolling. All together, the finite value of $t_{\text{lag}}$ is a remnant of a second order phase transition, with its large magnitude originating from the   singularity at the transition point determined by [8].
\\

\subsection*{Experiments}
To test our results on the scaling law, we used experiments with 3D-printed cylinders of irregular cross-section shapes and rolled them down an inclined ramp, with a fixed initial orientation for each set of runs (Figure \ref{fig:2D_Expt} a) (Materials and Methods). The ramp is coated with a layer of rubber and the adhesive force between the rubber and the cylinder leads to an effective viscous resistance to motion, the magnitude of which can depend on the rolling speed \cite{Maugis1990,Goohpattader2011}. We measure the time taken for the cylinder to roll over a small distance at various $\alpha$, and this time is taken to be the lag time $t_{\text{lag}}$. While doing so, we capture the rolling motion over a small portion of the cylinder circumference, and by decreasing the inclination $\alpha$ slowly we capture the singularity within that small portion. 5 runs are performed with varying mass and shape of the cylinder, and the singular behaviour of the measured $t_{\text{lag}}$ is apparent in the linear plot (Figure \ref{fig:2D_Expt} b). For each run we fit an $\alpha^{*}$ to the data, and a log-log plot of $t_{\text{lag}}$ against $\alpha-\alpha^{*}$ reveals the power-law scaling predicted by [8] (Figure \ref{fig:2D_Expt} c). See SI Movies S1 - S4 for representative videos of fast, medium, slow, and very slow rolling motion.
\\

\section*{3D rolling}

\subsection*{Equations of motion}
Moving to the case of the rolling of a sphere, we again consider torque and force balances, and the no-slip condition at the contact point which are now given by \cite{Levi2014}
\eqS
\begin{split}
I\dot{\bm\omega}&=\bm{r}\cross\bm{f}-\gamma\bm{\omega},\\
m\dot{\bm{v}}&=\bm{f}+\bm{F},\\
\bm{0}&=\bm{v}+\bm\omega\cross\bm{r}.
\end{split}
\eqE
Here $\bm{\omega}$ is the angular velocity vector in the lab frame, $I=\frac{2}{5}m\expval{r}^{2}$ is the average moment of inertia of the sphere of nominal radius $\expval{r}$, $\bm{v}$ is the velocity of the center of mass of the sphere, and $\bm f$ and $\bm F$ are the external forces due to contact and gravity respectively. These can be combined to give a single equation of motion (SI section 2A) for the rolling of the spheroid:
\eqS
\begin{split}
\dot{\bm\omega}=&-\frac{1}{I+mr^{2}}(\bm{r}\cross\bm{F}+\gamma\bm\omega)\\
&-\frac{1}{I+mr^{2}}\frac{m\gamma}{I}(\bm{r}\cdot\bm\omega)\bm{r}-\frac{m}{I+mr^{2}}\bm{r}\cross(\bm\omega\cross\dot{\bm{r}}).
\end{split}
\eqE
In the limit of overdamped viscous rolling (SI section 2A), the equation simplifies to
\eqS
\bm\omega=-\frac{1}{\gamma}\bm{r}\cross\bm{F}.
\eqE

\begin{figure}[htb]
\includegraphics[width=8cm]{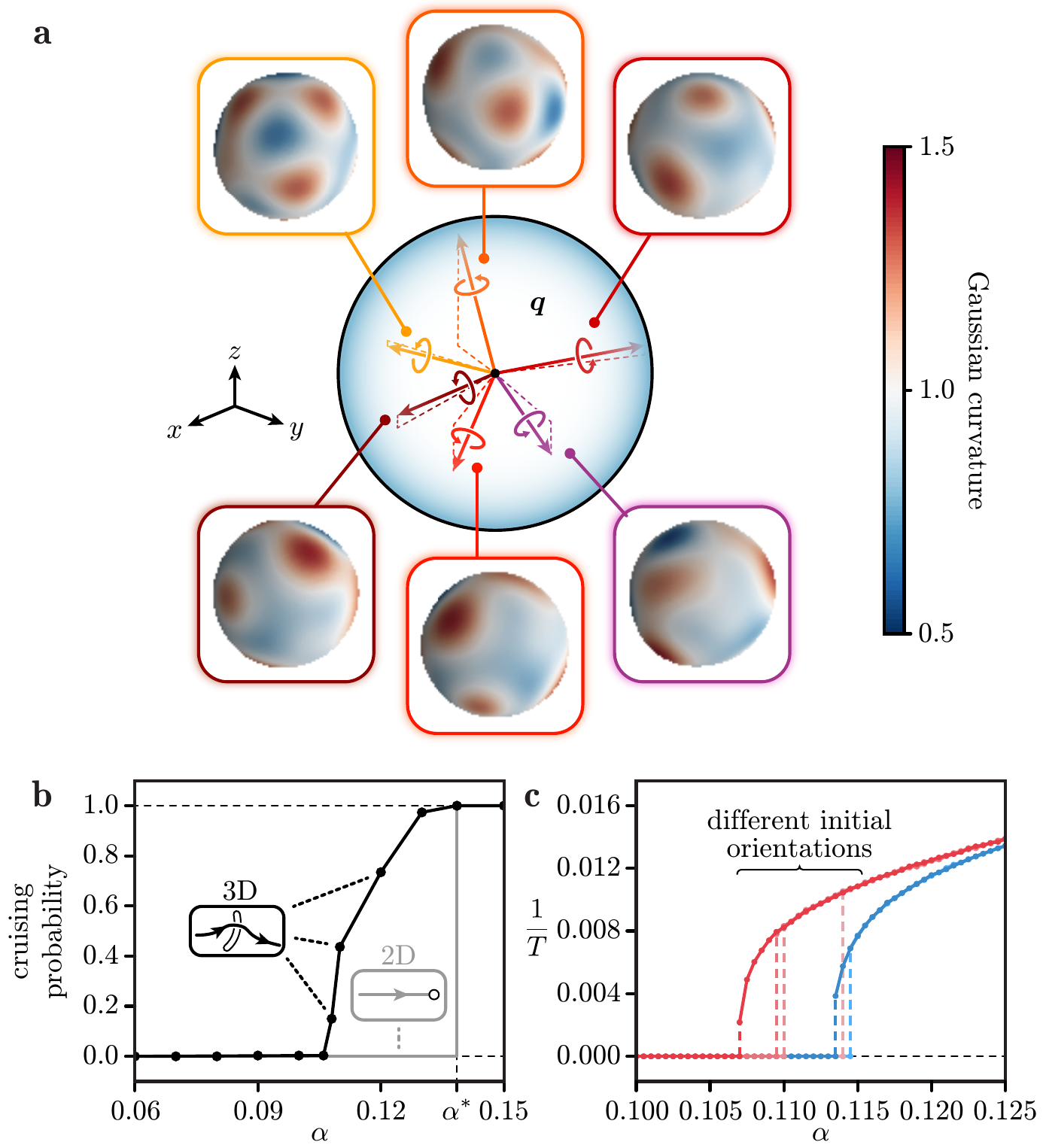}
\caption{\textbf{Rolling in 3D exhibits first-order phase transition characters in the viscous regime.} \textbf{a} In the quaternion representation, each vector $\bm{q}$ in the unit 3-ball (centre sphere) encodes a 3D orientation of the rolling sphere (peripheral panels, colours correspond to the local Gaussian curvature). \textbf{b} Simulation results of the same sphere released from 300 different initial orientations at a range of $\alpha$. Since halting states and cruising states can co-exist, some spheres could fall into the cruising state even when $\alpha<\alpha^{*}$ and the proportion of cruising spheres increases as $\alpha$ approaches $\alpha^{*}$, reaching 1 at $\alpha=\alpha^{*}$ (black). In contrast, halting and cruising states cannot co-exist in 2D rolling because of the dimensionality constraint, and a step function is expected for cruising probability with the same $\alpha^{*}$ (grey). Inset: a cruising trajectory (solid lines) can circumvent the halting state (hollow line and hollow point) using the extra dimension in 3D, while this is not possible in 2D. \textbf{c} $1/T$ values for the same rolling sphere with 6 different initial orientations as $\alpha$ is increased. Dashed lines correspond to discontinuous jumps from $\frac{1}{T}=0$ to some non-zero rolling speed. Blue and red colours correspond to 2 cruising orbits that the sphere converges to at large $\alpha$, identified by examining trajectories of individual simulation runs.}
\label{fig:3D_PT}
\end{figure}

Similar to our analysis of 2D rolling, the presence or absence of halting states  is controlled by the term $\bm{r}\cross\bm{F}$, suggesting that we need an expression for $\bm{r}$.  In the 3D body frame, the contact and normal vectors have components (Figure \ref{fig:setup} e)
\eqS
\begin{split}
\bm{r}_{0}=r(\theta,\phi)\cdot\begin{pmatrix}\sin\theta\cos\phi\\\sin\theta\sin\phi\\\cos\theta\end{pmatrix},\quad\bm{n}_{0}=\frac{\partial_{\theta}\bm{r}_{0}\cross\partial_{\phi}\bm{r}_{0}}{|\partial_{\theta}\bm{r}_{0}\cross\partial_{\phi}\bm{r}_{0}|}.
\end{split}
\eqE
Here we have used the shorthand $\partial_{\theta}\equiv\frac{\partial}{\partial\theta}$ and $\partial_{\phi}\equiv\frac{\partial}{\partial\phi}$.To compute their lab frame counterparts, we use a quaternion representation which does not suffer the coordinate singularities associated with Euler angles, and write
\eqS
\begin{split}
\bm{r}=q\bm{r}_{0}\bar{q},\quad\bm{n}=q\bm{n}_{0}\bar{q},\\
\end{split}
\eqE
where $q$ and $\bar{q}$ are the unit quaternion and its conjugate \cite{Sola2017,Altmann2005} and encode 3D orientation information with a form given by
\eqS
q=\cos\frac{\psi}{2}+\sin\frac{\psi}{2}\bm{u},\quad\bar{q}=\cos\frac{\psi}{2}-\sin\frac{\psi}{2}\bm{u},
\eqE
where $\bm{u}$ is a unit vector in lab frame around which the object is rotated by angle $\psi$ (see SI section 2B and \cite{Sola2017,Altmann2005} for further details on quaternion algebra). We note that the vector part $\bm{q}\equiv\sin\frac{\psi}{2}\bm{u}$ lives in a unit 3-ball with each $\bm{q}$ representing a 3D orientation (Figure \ref{fig:3D_PT} a) and the equation of motion for $q$ is simply
\eqS
\dot{q}=\frac{1}{2}\bm{\omega}q.
\eqE
\\
Here, we pause to distinguish and contrast 3D rolling with the 2D case. The no-slip condition in 2D is integrable (holonomic) and allows us to establish an explicit relation between the rotation angle $\psi$ and contact angle $\theta$ through Equation (2) to eliminate $\psi$. This allows us to retain $\theta$ as the state variable because the radius $r(\theta)$ is most conveniently characterised using $\theta$. 
The 3D rolling condition is not integrable (non-holonomic) \cite{Bloch2005,Bloch2015} and couples the orientation and angular velocity via the condition of no-slip. More explicitly, retaining $\theta$ and $\phi$ as the only state variables is insufficient since three quantities are needed to fully characterize the body orientation. 

To make progress, we keep $q$ as well as $\theta$ and $\phi$ as state variables in numerical simulations.
Computing $\bm\omega$ and $\dot{\bm\omega}$ allows the quaternion $q$ to be propagated, while we still need to update the contact angles. This is done by invoking the contact condition that the lab frame normal vector is always perpendicular to the plane, leading to
\eqS
\dot{\bm{n}}=\bm\omega\cross\bm{n}+\begin{pmatrix}\partial_{\theta}\bm{n}&\partial_{\phi}\bm{n}\end{pmatrix}\begin{pmatrix}\dot{\theta}\\\dot{\phi}\end{pmatrix}=\bm{0}.
\eqE
After some algebraic manipulations this allows to arrive at explicit expressions for $\dot\theta$ and $\dot\phi$ (see SI section 2C for details).


\subsection*{Halting-cruising transitions}

The halting condition can be worked out just as in the 2D case, since an equilibrium orientation corresponds to the position vector $\bm{r}$ being parallel to $\bm{F}$. It is then easy to deduce that a stable equilibrium state cannot exist if $\frac{\bm{r}}{r}\cdot\bm{n}\leq\cos\alpha$. To re-cast it to the form similar to the 2D case, we write the critical $\alpha^{*}$ as (SI section 2D)
\eqS
\alpha^{*}=\tan^{-1}\left[\max_{\theta,\phi}|\bm{\nabla}_{s}\ln r(\theta,\phi)|\right].
\eqE
$\bm{\nabla}_{s}$ is the spherical part of the gradient operator, and $|\bm{\nabla}_{s}r|=\sqrt{(\partial_{\theta}r)^{2}+(\partial_{\phi}r)^{2}/\sin^{2}\theta}$. When $\alpha<\alpha^{*}$ the halting state corresponds to sets of points where the ball is completely stationary, and these sets are 1D loops (see SI section 2E for details of these states and their stability). Given the spatial and multiply-connected nature of $SO(3)$, stable 1D cruising trajectories can co-exist with 1D halting loops, even in the absence of inertia. Consequently, we expect a first order transition in rolling speed to occur, and no stationary state should exist when $\alpha>\alpha^{*}$.

\begin{figure*}[htb]
\includegraphics[width=16cm]{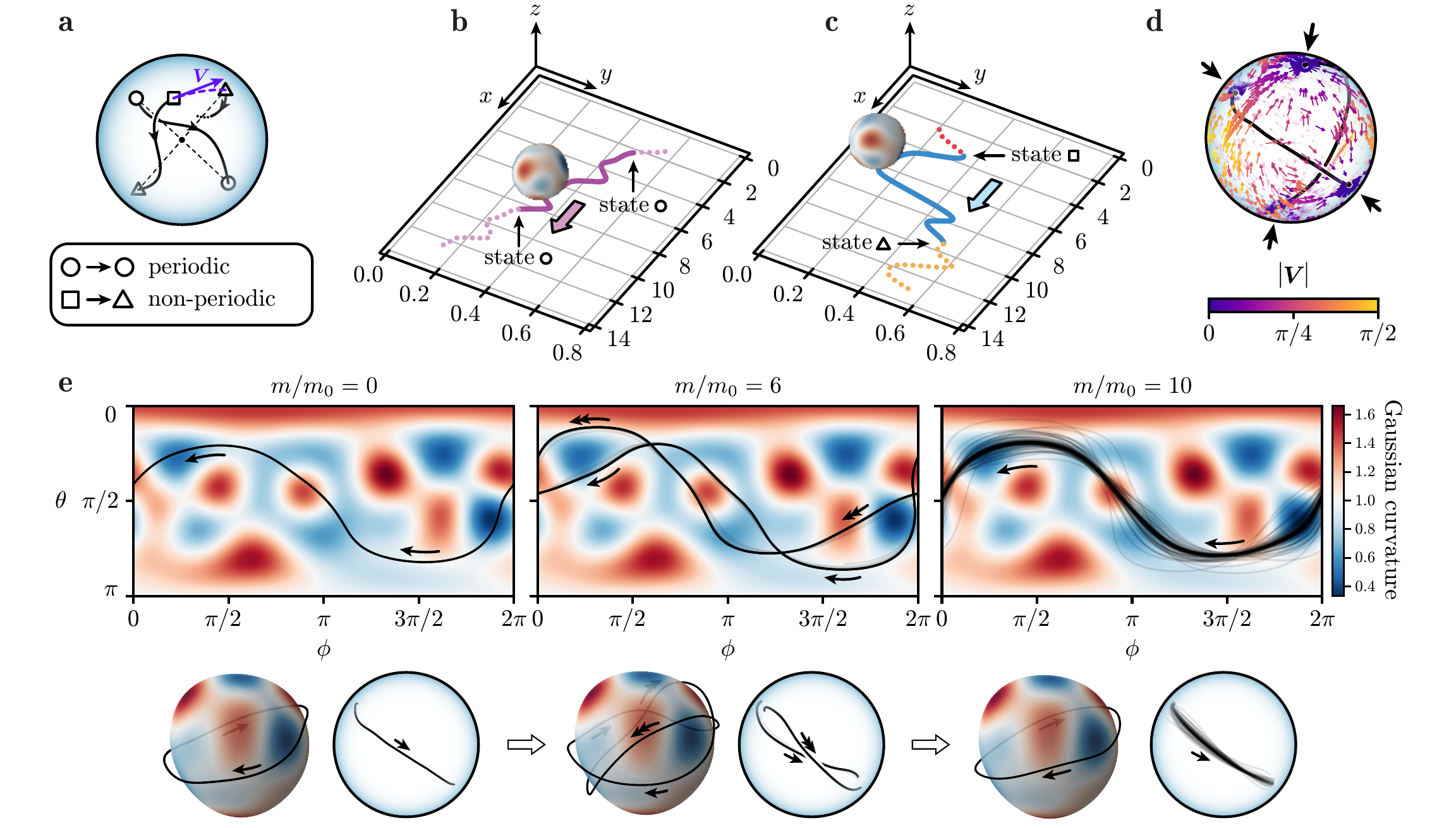}
\caption{\textbf{Periodicity and period-doubling of 3D rolling.} \textbf{a} In the viscous regime, a Poincar\'e map in $SO(3)$ elucidates emergence of periodic orbits. To construct the map, we place initial states on $\partial\Omega$ (circle and square markers) and evolve them (solid lines) until they reach $\partial\Omega$ again, and perform an antipodal identification (dashed lines). A closed segment is formed if the antipodal point of the final state is the same as the initial state (circle markers) and the spatial trajectory will be periodic. On the other hand, if a state on $\partial\Omega$ reaches a point that is not antipodal (square to triangle) then we can define a vector $\bm{V}$ that points from the initial to the final state after one antipodal identification (purple arrow), thus arriving at a Poincar\'e map. The spatial trajectory of this open segment is not periodic. \textbf{b}, \textbf{c} Simulation results of spatial trajectories formed by a closed and an open segment respectively. Colours on the ball represent Gaussian curvature values. \textbf{d} We sample the vector field $\bm{V}$ on $\partial\Omega$ through numerical simulations. Here $\alpha>\alpha^{*}$ so no halting state exists in $\Omega$. 2 pairs of attractors are seen (black arrows), corresponding to 2 distinct stable, periodic orbits (solid lines going through the sphere). Arrow colors correspond to the length $|\bm{V}|$. \textbf{e} When inertia is added to the system, period-doubling is observed. Top panels: plots of late-time $[\theta(t),\phi(t)]$ for 100 simulations (black lines). The period-1 orbits at $m/m_{0}=0$ (left) become period-2 at $m/m_{0}=6$ (middle), and tend to the ideal orbit when mass is increased further (right). Fluctuations in the $m/m_{0}=10$ case originate from a long equilibration time. Bottom panels: illustrations of wrapping of the contact point around the sphere and corresponding rolling orbits in $SO(3)$. Period doubling at $m/m_{0}=6$ is characterised by the initial closed orbit splitting in two.}
\label{fig:3D_Comp}
\end{figure*}

We verify this using numerical simulations of overdamped (viscous) rolling of a fixed shape with different random initial orientations (SI section 2F). As expected, the sphere remains stationary for small $\alpha$, but can start rolling/cruising when $\alpha$ increases, but the actual value of this critical angle depends on its initial state. On the ensemble level, the proportion of initial orientations that enter the cruising state increases as $\alpha$ increases, since fewer and fewer halting states can exist for larger $\alpha$ (Figure \ref{fig:3D_PT} b). For individual simulation traces, discontinuous jumps are observed to different equilibrium rolling speeds, occurring at different $\alpha$ depending on their initial orientations (Figure \ref{fig:3D_PT} c), reflecting the fact that they start out being trapped in different halting states. However, we find that the equilibrium cruising trajectories are all periodic. Computing the period $T$ for each initialization as a function of $\alpha$ allows us to consider the frequency $\frac{1}{T}$  as an order parameter. In Figure \ref{fig:3D_PT} c, we show that  when the $\frac{1}{T}-\alpha$ curves are plotted together for different initial orientations, the data collapse onto one of 2 universal curves representing 2 stable periodic orbits.
\\

\subsection*{Topological signatures in 3D viscous rolling}

\begin{figure*}[htb]
\includegraphics[width=16cm]{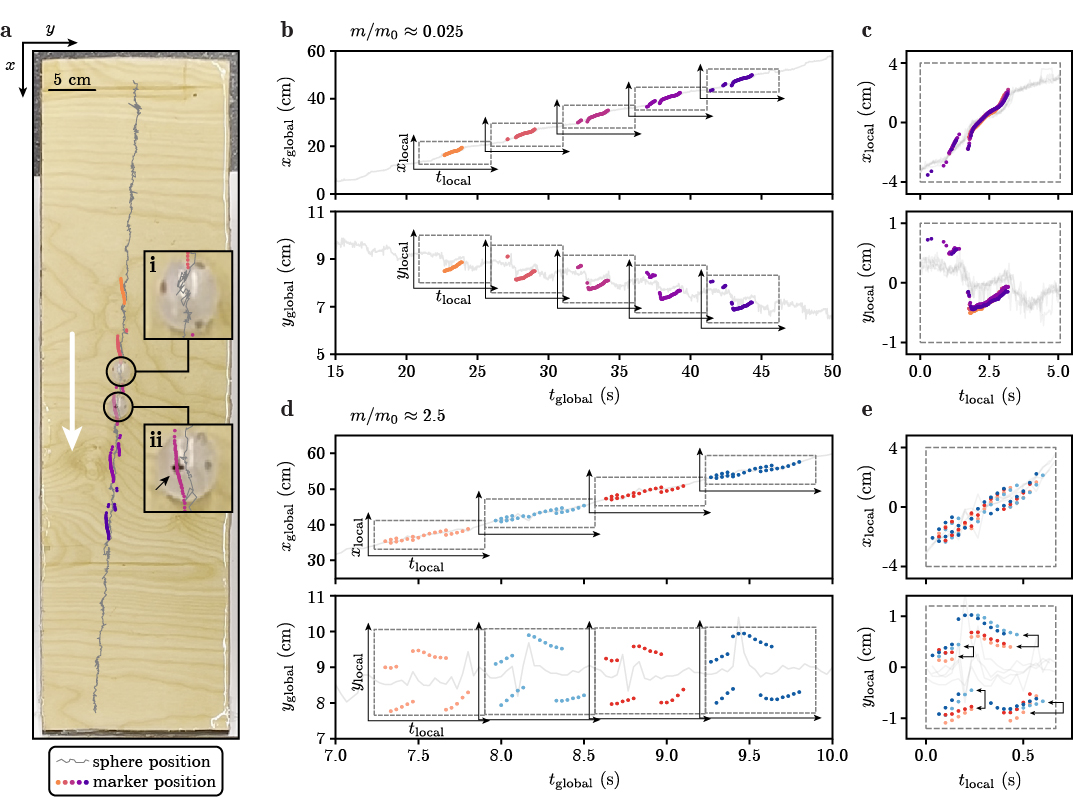}
\caption{\textbf{Experimental verification of 3D viscous rolling's periodic nature and period-doubling in inertial rolling.} \textbf{a} A 3D-printed irregular sphere is rolled down a ramp, and the position of the sphere is tracked (grey solid line). Black markers on the sphere are traced at the same time (coloured scatter points). Inset: \textbf{i} close-up snapshot when markers are not detected and \textbf{ii} when one marker is dark enough to be seen (black arrow).  \textbf{b} For viscous rolling, both the sphere and the ramp is coated with a rubber substrate. We estimate $m/m_{0}\approx0.025$. Time traces of $x_{\text{global}}$ (top) and $y_{\text{global}}$ (bottom) coordinates of the sphere (grey solid lines) and markers (scatter points) are plotted as a function of $t_{\text{global}}$. The local coordinates $x_{\text{local}}$, $y_{\text{local}}$, and $t_{\text{local}}$ can be constructed by identifying self-similar patterns in the marker positions (axes within plots). \textbf{c} When plotted in local frames, the sphere trajectories (grey lines) and markers (coloured scatter points) show significant overlay, indicating the motion is periodic. The sphere has an average radius of 1 cm, leading to an $x_{\text{local}}$ extent of about 6 cm. \textbf{d} Without rubber coating on the sphere, we estimate $m/m_{0}\approx2.5$. Alternating blue and red are used to colour marker positions for four consecutive rolling groups. \textbf{e} When plotted in local coordinates, the $y_{\text{local}}$ marker positions exhibit period-doubling characters. The pair of blue-coloured periods have marker positions resembling each other and the same can be said for the pair of red periods. However, comparing blue to red, two distinct trends can be identified (paired arrows), indicating that the trajectory is now self-repeating only after the sphere has rolled over itself twice. Note the extent of $x_{\text{local}}$ is identical to that in \textbf{c}.}
\label{fig:3D_Expt}
\end{figure*}

Why is the viscous rolling motion periodic, instead of mimicking a random walk? To trace the origin of these 2 periodic cruising states, recall that an orientation state can be represented by a point in the unit 3-ball $\Omega$, and a Poincar\'e map $\bm{V}$ can be constructed on the surface $\partial\Omega$ by computing the difference between an initial state starting on $\partial\Omega$, and the final state when it traverses the interior of $\Omega$ and hits $\partial\Omega$ again, after an antipodal identification (Figure \ref{fig:3D_Comp}a, b, and c). The vector field $\bm{V}$ is continuous and single-valued on the 2-sphere $\partial\Omega$ so by the Ulam-Borsuk (hairy ball) theorem, which states that \textit{there is no non-vanishing continuous tangent vector field on even-dimensional n-spheres}, there must exist points on $\partial\Omega$ where $\bm{V}=\bm{0}$, which correspond to closed periodic cruising orbits. 
\\

In the case of rolling, the existence of closed orbits on its own does not directly translate into periodic rolling: these orbits also need to be stable. By reversing the force $\bm{F}\to-\bm{F}$ while still keeping the non-slip constraint, we can construct a time-reversed system where a stable orbit becomes unstable and vice versa. 
Physically, time-reversal in our system corresponds to rolling of a sphere stuck to a ceiling tilted at angle $\alpha$. It is not unreasonable to assume that this motion should not be fundamentally too different from the normal downward rolling, and as such we expect stable and unstable orbits to both exist. To visualise the stable rolling orbits, we construct $\bm{V}$ explicitly from numerical simulations and found there are 4 attractor points on $\partial\Omega$, corresponding to two pairs of antipodal points with $\bm{V}=\bm{0}$ and thus 2 closed stable orbits as seen earlier (Figure \ref{fig:3D_Comp} d). Why are there 2 closed orbits instead of any other number? Intuitively, every 3D object has a long axis and this axis will most likely be parallel to the floor and perpendicular to the sloping direction in a cruising state, and there are 2 ways to orient this axis, leading to 2 distinct trajectories. When illustrated in $SO(3)$, one of the 2 orbits appears to run through the middle of the 3-ball while the other moves close to the boundary and forms a semi-circular shape (Figure \ref{fig:3D_Comp} d). To make sense of this arrangement, we compute rolling trajectories of a perfect sphere and highlight those that pass through an arbitrary contact point, thus visualising the effect of a point perturbation on perfect rolling (SI section 2G). The set of highlighted states forms a tube running though the middle of $SO(3)$ (Figure S3), sectioning the space into 2 parts, and the shapes of stable orbits follow closely the shapes of each sub-section. The non-trivial topology of $SO(3)$ compared to $SO(2)$ thus introduces nuances in cruising states but the core phenomenology of first order transitions remain.
\\

\subsection*{Inertial effects in 3D rolling}
In the presence of inertia, generic behaviours of 3D rolling do not differ qualitatively from those in 2D - closed, stable orbits still exist, although the time taken for the rolling trajectory to converge to the stable orbits is significantly longer due to inertia. Interestingly, computing the rolling trajectories at intermediate $m/m_{0}$, where $m_{0}$ is the mass scale as defined before in Equation (4), reveals a period-doubling behaviour. We simulate rolling of the same random sphere from 100 different initial orientations at $\alpha=0.15>\alpha^{*}$, and plot $[\theta(t),\phi(t)]$ for $600t_{0}<t<700t_{0}$ (Figure \ref{fig:3D_Comp} e). In the viscous regime with $m/m_{0}=0$, the orbiting period is around 60 time units and the $[\theta(t),\phi(t)]$ curve traverses the $\theta-\phi$ space once before closing (Figure \ref{fig:3D_Comp} e, left). At intermediate $m/m_{0}=6$, the contact point wraps around the sphere twice before closing instead (Figure \ref{fig:3D_Comp} e, middle). At even larger $m/m_{0}=10$, convergence is much slower but no clear period-doubling is observed (Figure \ref{fig:3D_Comp} e, right). The trajectories plotted all correspond to rolling motions where $\phi(t)$ decreases, and depending on the initial orientation some trajectories have increasing $\phi(t)$ instead. This second set originates from the other closed orbit and their behaviours are qualitatively similar to the first set of trajectories (not shown in the figure).
\\

To make sense of the period-doubling phenomenon, recall that $SO(3)$ is doubly-connected, so a period-1 orbit does not necessarily bring the rolling body back to its initial state. With $m/m_{0}=0$, the hairy ball theorem constrains the orbit to be period-1 as explained before. However, when $m$ is non-zero, the hairy ball theorem is lifted and only a period-2 orbit brings the body back to the starting state. As $m$ increases further, the effect of surface irregularity diminishes and the orbit should approach that of a perfect sphere, which is again a period-1 orbit in $SO(3)$, a trend that is indeed observed in simulations (Figure \ref{fig:3D_Comp}e, right). This period-doubling of the orbit at intermediate $m$ is thus a manifestation of the doubly-connected nature of $SO(3)$ in a classical system.
\\

\subsection*{Experiments}
To test our predictions on the rolling periodicity, we 3D-printed irregular spheres and use the same experimental setup as in 2D rolling (Figure  \ref{fig:2D_Expt} a, \ref{fig:3D_Expt} a). For viscous rolling, we coat the sphere with rubber to increase the adhesive drag. To track the rolling trajectory, black markers are drawn on the sphere and we analysed videos of the sphere rolling down by quantifying both the position of the sphere and positions of the markers over time (Figure \ref{fig:3D_Expt} a, SI Movie S5). The sphere position allows us to automatically zoom-in onto the sphere (Figure \ref{fig:3D_Expt} a, insets), and marker positions serve as more accurate and precise indicators of the orientation information over time. The extracted spatial and temporal coordinates are denoted $x_{\text{global}}$, $y_{\text{global}}$, and $t_{\text{global}}$ (Figure \ref{fig:3D_Expt} b). To uncover the rolling periodicity, we introduce local coordinates by first identifying time windows where the trajectories appear to have repeating patterns, and $t_{\text{local}}$ can be defined by shifting all time coordinates in a window by a constant amount. In the second step, spatial coordinates are shifted in a similar fashion to produce $x_{\text{local}}$ and $y_{\text{local}}$ (Figure \ref{fig:3D_Expt} b, axis within plots). The shift parameters are chosen such that when plotted in local coordinates, the tracked trajectories and markers coincide (Figure \ref{fig:3D_Expt} c). This coincidence of tracked points demonstrates the periodic nature of 3D viscous rolling. In addition, using the rolling speed in the $x$-direction $v_{x}\approx10^{-2}$ ms$^{-1}$, we estimate an effective mass of $m/m_{0}=\frac{v_{x}^{2}}{g\expval{r}\sin\alpha}\approx0.025$ with $g$ the gravitational acceleration (SI section 2H), so the viscous regime approximation is valid. 

To study inertial rolling, we omit the coating on the sphere and perform the same experiments as before. The resulting rolling speed increases to $v_{x}\approx10^{-1}$ ms$^{-1}$, leading to a reduced mass of approximately 2.5 (SI section 2H, SI Movie S6). We again track the position of the sphere and positions of markers (Figure \ref{fig:3D_Expt} d), and plot them in the shifted, local coordinate system (Figure \ref{fig:3D_Expt} e). We plot the marker points' positions using alternating blue and red colours for each period, and see the existence of period-doubling by observing that in the $y_{\text{local}}$ coordinate, marker points of the same colour form similar patterns, but those of different colours form different patterns (Figure \ref{fig:3D_Expt} e, paired arrows in bottom panel). We interpret this observations by noting that this follows from the doubly-connected nature of $SO(3)$.

\section*{Discussion}

Rolling appears to be a mundane, everyday process but it connects a surprisingly diverse range of ideas, ranging from the role of non-holonomic constraints,  stochasticity in shape and forcing, and the nontrivial nature of 3D rotations, all deeply rooted in geometry and physics.

Our analysis of the rolling dynamics of 2D and 3D objects with randomness in their shapes shows the appearance of first and second order phase transitions in terms of the terminal rolling speed treated as an order parameter. Furthermore, our analysis of 3D rolling behaviours allows us to see dynamical manifestations of mathematical theorems associated with the nature of vector fields on spheres, and the topology of the 3d rotation group, e.g. period-1 and period-2 rolling orbits. 

While our initial motivation for this study was mere curiosity, perhaps our conceptual framework  can potentially be applied to a range of other fields including robotics and biology, where effects of shape randomness on transport driven by rolling do not seem to have been extensively explored. For example, in robotics, by controlling the softness of a rolling robot, it is possible to control its shape under the influence of gravity and thence its trajectory \cite{Yanagida2017,Hatton2010,Gong2012,Sobolev2023}. This begs the question of optimal control of sphere softness to guide a robot similar to BB-8 in the Star Wars saga \cite{Burkhardt2016,Ilin2017}. A natural manifestation of this might be seen in the shape and trajectory of dung balls rolled by the eponymous dung beetle \cite{Dacke2021,Tomkins1999}: the dung ball is an irregular spheroidal object guided by an external agent, raising the question of how to best do it. In biology on the cellular scale, the interplay between rolling and thermal noise in appears naturally in the context of active vesicle transport \cite{Zheng2023a,Liao2019,Verdeny-Vilanova2017}.  Our study suggests the need to look more carefully at the effect of size and shape of vesicles on the ability of molecular motors to move them, especially in the context of the noise amplifying effect of singularities close to criticality in the driving force, the active analog of $\alpha$, in both natural and artificial settings, all problems for the future.

\textbf{Materials and methods.} The 2D experiment is done by rolling a 3D-printed cylinder on a rubber-coated wooden plank. The coating used is the Ecoflex 00-30 Platinum Cure Silicon Rubber Compound, and to avoid noise effects arising from spatial inhomogeneity of the coating we restrict the rolling motion to a fixed small region on the ramp. Two blocks are placed on the plank that mark the start and end positions, and in addition a mark is made on the cylinder and always aligned perpendicular to the plank at the starting position. This ensures the starting orientation is kept constant. 
\\

To allow for precise adjustment of $\alpha$, a 72.2-cm long plank is used. One end of the plank rests on the bench and the other end is lifted and placed onto an adjustable frame, where the height of the frame is changed using a micro-meter screw gauge. Close to the singularity, the vertical displacement between points in one run is varied by as low as 0.25mm. This enables us to probe the change in $t_{\text{lag}}$ over very small changes in $\alpha$.
\\

The measurement of $t_{\text{lag}}$ in each run, for each $\alpha$, is done by manually placing the cylinder in the same position and orientation, and then releasing the cylinder from rest. A timer starts as it is released, and ends when the cylinder hits the end marker. 5 measurements are made for each $\alpha$, and the magnitude of change in $\alpha$ is reduced as $t_{\text{lag}}$ approaches the singularity. The precise value of $\alpha^{*}$ is not known \textit{a priori} since in practice, the cylinder is in contact with the ramp not at a single point but over a region due to deformation of the rubber substrate. 
\\

\textbf{Author contributions.} L.M. conceived of the study. D.Q.~and L.M.~designed the study. D.Q.~performed theoretical and computational investigations, D.Q.~and Y.J.~performed experiments. All authors wrote, reviewed and edited the manuscript.
\\

\textbf{Acknowledgement.} The study is funded by Transition Bio Ltd (D.Q.), the National Research Foundation of Korea
(Grant No. 2021R1A6A3A03039239) (Y.J.), the Simons Foundation (L.M.) and the Henri Seydoux Fund (L.M.). The authors thank Tuomas P.~J.~Knowles, ~Ismael Sierra, and ~Puskar Mondal for their support and helpful discussions. All authors declare no competing interest.
\\

\bibliography{library}

\end{document}


\maketitle


\section{2D Rolling}
\subsection{Equation of motion}
\label{2DEOM}
In 2D, the torque and force balance equations are
\eqS
\begin{split}
I\ddot{\psi}&=(\bm{r}\cross\bm{f})\cdot\bm{\hat{k}}-\gamma\dot{\psi}\\
m\dot{\bm{v}}&=\bm{f}+\bm{F}
\end{split}
\eqE 
and the aim is to write the equation of motion, obtained by combining the two above,
\eqS
\begin{split}
I\ddot{\psi}&=\left[\bm{r}\cross\left(m\dot{\bm{v}}-\bm{F}\right)\right]\cdot\bm{\hat{k}}-\gamma\dot{\psi}
\end{split}
\eqE 
in terms of $\theta$, $r(\theta)$ and their derivatives alone. We proceed by first writing down the form of the vectors from definition,
\eqS
\begin{split}
\bm{r}=&r(\theta)\cdot\begin{bmatrix}\cos(\psi+\theta)\\\sin(\psi+\theta)\\0\end{bmatrix}\\
\bm{\hat{k}}=&\begin{bmatrix}0\\0\\1\end{bmatrix}\\
\bm{F}=&F\cdot\begin{bmatrix}\sin\alpha\\-\cos\alpha\\0\end{bmatrix}\\
\bm{v}=&\dot{\psi}\bm{r}\cross\bm{\hat{k}}=\dot{\psi}r(\theta)\begin{bmatrix}\sin(\psi+\theta)\\-\cos(\psi+\theta)\\0\end{bmatrix}
\end{split}
\eqE
where the last line comes from the non-slip condition $\bm{v}+\bm{r}\cross(-\dot{\psi}\bm{\hat{k}})=\bm{0}$. To deal with the sum $\psi+\theta$, we use $\Delta\theta$, which is the angular difference between the contact vector $\bm{r}_{0}$ and the unit normal $\bm{n}_{0}$, as well as the constraint that $\bm{n}$ points perpendicular into the plane to write
\eqS
\psi+\theta=\frac{3}{2}\pi+\Delta\theta
\eqE
and an expression for $\Delta\theta$ will allow us to relate $\psi$ to $\theta$. 
\\

\begin{figure}[htb]
\centering
\includegraphics[width=5in]{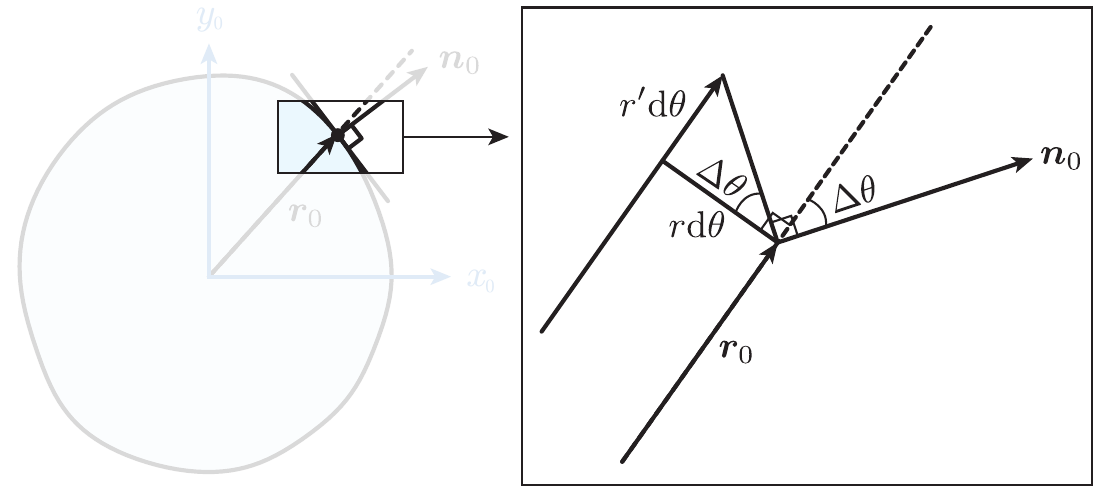}
\caption{Expressing $\Delta\theta$ as a function of $r(\theta)$. The infinitesimal angle $\dd\theta$ allows us to construct line elements $r\dd\theta$ and $r'\dd\theta$ that forms the triangle relating $\Delta\theta$ to $r/r'$.}
\label{fig:SI_2D}
\end{figure}

Since the normal vector is perpendicular to the line element on the disk's perimeter, direct inspection (Fig \ref{fig:SI_2D}) gives the relation
\eqS
\tan\Delta\theta=\frac{r'}{r}
\eqE
where $r'=\frac{\dd r}{\dd\theta}$. We can now express all variables as a function of only $\theta$. Some helpful expressions are
\eqS
\begin{split}
\tan(\psi+\theta)=&-\cot\Delta\theta=-\frac{r}{r'}\\
\sin(\psi+\theta)=&-\frac{r}{\sqrt{r^{2}+r^{\prime2}}}\\
\cos(\psi+\theta)=&+\frac{r'}{\sqrt{r^{2}+r^{\prime2}}}\\
\frac{\dd \psi}{\dd \theta}=&\frac{\dd}{\dd\theta}(\Delta\theta-\theta)=\frac{r''/r-r^{\prime2}/r^{2}}{1+\left(r'/r\right)^{2}}-1\\
=&-\sqrt{r^{2}+r^{\prime2}}\kappa
\end{split}
\eqE
where $\kappa=\frac{r^{2}+2r^{\prime2}-rr''}{\sqrt{r^{2}+r^{\prime2}}^{3}}$ is the curvature. The last relation allows us to write the time derivatives of $\psi$ as
\eqS
\begin{split}
\dot{\psi}=&-\sqrt{r^{2}+r^{\prime2}}\kappa\dot\theta\\
\ddot{\psi}=&-\frac{\dd}{\dd t}(\sqrt{r^{2}+r^{\prime2}}\kappa\dot\theta)\\
=&-\sqrt{r^{2}+r^{\prime2}}\kappa\ddot\theta-\left(\sqrt{r^{2}+r^{\prime2}}\kappa\right)'\dot\theta^{2}
\end{split}
\eqE
and the trigonometric relations give expressions for $\bm{r}$, $\bm{v}$ and $\dot{\bm{v}}$:
\eqS
\begin{split}
\bm{r}=&\frac{1}{\sqrt{r^{2}+r^{\prime2}}}\begin{pmatrix}rr'\\-r^{2}\\0\end{pmatrix}\\
\bm{v}=&\frac{\dot{\psi}}{\sqrt{r^{2}+r^{\prime2}}}\begin{pmatrix}-r^{2}\\-rr'\\0\end{pmatrix}=\kappa\begin{pmatrix}r^{2}\\rr'\\0\end{pmatrix}\dot\theta\\
\dot{\bm{v}}=&\begin{bmatrix}
\kappa r^{2}\ddot\theta+(\kappa r^{2})'\dot{\theta}^{2}\\
\kappa rr'\ddot\theta+(\kappa rr')'\dot{\theta}^{2}\\0
\end{bmatrix}.
\end{split}
\eqE
Substituting these into the equation of motion gives
\eqS
\ddot{\theta}=\frac{1}{I+mr^{2}}\left[
\frac{Fr (r\sin\alpha-r'\cos\alpha)}{\kappa \left(r^2+r^{\prime2}\right)}-\gamma  \dot\theta \right]-\left[\frac{I+mr^{\prime2}}{I+mr^{2}}r(r\kappa)'+\kappa r'r''+\frac{mr^{2}}{I+mr^{2}}\left(r^{2}\kappa\right)'+\frac{I}{I+mr^{2}}r^{\prime2}\kappa'\right]\frac{\dot{\theta}^{2}}{\kappa(r^{2}+r^{\prime2})}.
\eqE
We can write this in a dimensionless form using the time and mass scales
\eqS
\begin{split}
t_{0}\equiv&\frac{\gamma}{F\expval{r}},\\
m_{0}\equiv&\frac{\gamma^{2}}{F\expval{r}^{3}}.\\
\end{split}
\eqE
We use the following notations
\eqS
\begin{split}
    \ddot{\theta}&\equiv \frac{1}{t_{0}^{2}}\ddot{\tilde{\theta}},\\
    \dot{\theta}&\equiv \frac{1}{t_{0}}\dot{\tilde{\theta}},\\
    r&\equiv\expval{r}\tilde{r},\\
    \kappa&\equiv\frac{1}{\expval{r}}\tilde{\kappa},\\
\end{split}
\eqE
as well as $I=\frac{1}{2}m\expval{r}^{2}$, to give via direct substitution
\eqS
\begin{split}
\frac{1}{t_{0}^{2}}\ddot{\tilde\theta}=&\frac{F\expval{r}}{m\expval{r}^{2}}\frac{1}{\frac{1}{2}+\tilde r^{2}}\left[
\frac{\tilde r (\tilde r\sin\alpha-\tilde r'\cos\alpha)}{\tilde\kappa \left(\tilde r^2+\tilde r^{\prime2}\right)}-\dot{\tilde{\theta}} \right]-\frac{1}{t_{0}^{2}}\left[\frac{\frac{1}{2}+\tilde r^{\prime2}}{\frac{1}{2}+\tilde r^{2}}\tilde r(\tilde r\tilde \kappa)'+\tilde \kappa\tilde r'\tilde r''+\frac{\tilde r^{2}}{\frac{1}{2}+\tilde r^{2}}\left(\tilde r^{2}\tilde\kappa\right)'+\frac{\frac{1}{2}}{\frac{1}{2}+\tilde r^{2}}\tilde r^{\prime2}\tilde\kappa'\right]\frac{\dot{\tilde\theta}^{2}}{\tilde\kappa(\tilde r^{2}+\tilde r^{\prime2})}\\
\ddot{\tilde\theta}=&\frac{m_{0}}{m}\frac{1}{\frac{1}{2}+\tilde r^{2}}\left[
\frac{\tilde r (\tilde r\sin\alpha-\tilde r'\cos\alpha)}{\tilde\kappa \left(\tilde r^2+\tilde r^{\prime2}\right)}-\dot{\tilde{\theta}} \right]-\left[\frac{\frac{1}{2}+\tilde r^{\prime2}}{\frac{1}{2}+\tilde r^{2}}\tilde r(\tilde r\tilde \kappa)'+\tilde \kappa\tilde r'\tilde r''+\frac{\tilde r^{2}}{\frac{1}{2}+\tilde r^{2}}\left(\tilde r^{2}\tilde\kappa\right)'+\frac{\frac{1}{2}}{\frac{1}{2}+\tilde r^{2}}\tilde r^{\prime2}\tilde\kappa'\right]\frac{\dot{\tilde\theta}^{2}}{\tilde\kappa(\tilde r^{2}+\tilde r^{\prime2})}.
\end{split}
\eqE
and in simulations, we set the dimensions to be $F=\expval{r}=\gamma=1$.
\\

\subsection{Second order phase transition in viscous limit}
In the viscous limit, we have
\eqS
\dot\theta=\frac{F}{\gamma}\frac{r(r\sin\alpha-r'\cos\alpha)}{\left(r^{2}+r^{\prime2}\right)\kappa}
\eqE
To study this effect, we compute the time period $T$ that $\theta$ takes to increase by $2\pi$. It has the expression
\eqS
T=\int_{0}^{2\pi}\frac{\dd t}{\dd\theta}\dd\theta
\eqE
and in the trivial case of a uniform disk, we set $r=\expval{r}$ to be a constant, so $\kappa=1/\expval{r}$, and the corresponding period is $2\pi\frac{\gamma}{F}\frac{1}{\expval{r}\sin\alpha}$. In the general case, $T$ has the expression
\eqS
T=\frac{\gamma}{F\sin\alpha}\int_{0}^{2\pi}\left[\frac{\left(r^{2}+r^{\prime2}\right)\kappa}{r^2}\right]\left[\frac{1}{1-\frac{r'}{r}\cot\alpha}\right]\dd\theta
\eqE
and the first square bracket of the integrand is always positive, while the second bracket could contains a singularity if at some $\theta$, $r\leq r'\cot\alpha$. It has a simple physical interpretation: if at any contact point we have $r=r'\cot\alpha$ the force $\bm{F}$ and contact vector $\bm{r}$ are parallel, so the disk would remain stationary in the absence of inertia. We can define a `quality factor' $Q(\theta)$ as
\eqS
Q(\theta)\equiv\frac{r'(\theta)}{r(\theta)}=\frac{\dd}{\dd\theta}\ln r(\theta)
\eqE
so the condition for the existence of a stationary state can be written as
\eqS
\max_{\theta}Q(\theta)\equiv Q^{*}\geq\tan\alpha
\eqE
and we use $\theta^{*}$ to denote the contact angle where $Q(\theta)$ attains the maximum value: $Q(\theta^{*})=Q^{*}$. Since the average of $Q(\theta)$ over all $\theta$ is 0 (it is a total differential), $Q(\theta)$ almost always crosses $\tan\alpha$ at pairs of points, one stable and one unstable.
\\

Criticality happens where for a given $Q^{*}$, the inclination angle $\alpha$ is exactly $\tan^{-1}Q^{*}\equiv\alpha^{*}$. Expanding $Q(\theta)$ near $\theta^{*}$, we write
\eqS
Q(\theta^{*}+\delta\theta)=\tan\alpha^{*}+Q''(\theta^{*})\delta\theta^{2}
\eqE
where $Q''(\theta^{*})<0$ because it is a maxima, and the value of $T$ is dominated by the integral over this singularity:
\eqS
\begin{split}
T\approx&\frac{\gamma}{F\sin\alpha^{*}}\left.\left[\frac{\left(r^{2}+r^{\prime2}\right)\kappa}{r^2}\right]\right|_{\theta=\theta^{*}}\int_{-\epsilon}^{+\epsilon}\frac{1}{-Q''(\theta^{*})\cot\alpha^{*}\delta\theta^{2}}\dd\delta\theta
\end{split}
\eqE
for some finite positive $\epsilon$, which leads to a diverging integral. When the angle $\alpha$ is slightly above $\alpha^{*}$, we write $\alpha=\alpha^{*}+\delta\alpha$ and
\eqS
\cot(\alpha^{*}+\delta\alpha)=\cot\alpha^{*}-\frac{\delta\alpha}{\sin^{2}\alpha^{*}}+\mathcal{O}(\delta\alpha^{2})
\eqE
so that the divergent part becomes
\eqS
\begin{split}
1-\frac{r'}{r}\cot\alpha\approx&1-\left[\tan\alpha^{*}+Q''(\theta^{*})\delta\theta^{2}\right]\left[\cot\alpha^{*}-\frac{\delta\alpha}{\sin^{2}\alpha^{*}}\right]\\
\approx&-Q''(\theta^{*})\cot\alpha^{*}\delta\theta^{2}+\frac{\delta\alpha}{\sin\alpha^{*}\cos\alpha^{*}}.
\end{split}
\eqE
The $\delta\alpha$ term removes the singularity, and the scaling of $T$ with $\delta\alpha$ can be worked out using the formula
\eqS
\int_{-\epsilon}^{+\epsilon}\frac{1}{C_{1}\delta\theta^{2}+C_{2}\delta\alpha}\dd\delta\theta=\frac{2\tan^{-1}\left(\frac{\sqrt{C_{1}}\epsilon}{\sqrt{C_{2}\delta\alpha}}\right)}{\sqrt{C_{1}C_{2}}}\frac{1}{\sqrt{\delta\alpha}}
\eqE
for constants $C_{1}$, $C_{2}$. Close to $\delta\alpha=0$ the numerator is simply $\frac{\pi}{2}$, so that
\eqS
T\approx\frac{T^{*}}{\sqrt{\delta\alpha}}
\eqE
where
\eqS
T^{*}\equiv\frac{\gamma}{F}\left.\left[\frac{\left(r^{2}+r^{\prime2}\right)\kappa}{r^2}\right]\right|_{\theta=\theta^{*}}\frac{\pi}{\sqrt{-Q''(\theta^{*})}}
\eqE
and the angular velocity $\frac{2\pi}{T}$ exhibits a second order phase transition
\eqS
\frac{2\pi}{T}\propto\begin{cases}0\quad&\text{if}\quad\alpha\leq\alpha^{*}\\\sqrt{\delta\alpha}\quad&\text{if}\quad\alpha>\alpha^{*}\end{cases}
\eqE
for small $\delta\alpha$.
\\

\subsection{Numerical simulations}
In simulations we set the physical parameters to unity $F=\gamma=\expval{r}=1$, collectively these determine dimensions of time, mass and length. The moment of inertia is set to $I=\frac{1}{2}m\expval{r}^2$, which is the expression of an isotropic disk. Generalising $I$ to different geometries is equivalent to changing a pre-factor and does not bring qualitative changes to the kinematics. We generate the 2D disk by writing $r(\theta)$ as a Fourier expansion
\eqS
r(\theta)=1+\sum_{l=2}^{M}\frac{A_{l}}{l^{2}}\sin\left[l(\theta-\theta_{l})\right].
\eqE
The coefficients $A_{l}$ are drawn from a normal distribution with zero mean and standard deviation $\sigma$: $A_{l}\sim\mathcal{N}(0,\sigma)$, and the phases $\theta_{l}$ are random variables uniformly distributed in $[0,2\pi)$. $M$ is the number of modes used. The $\frac{1}{l^{2}}$ factor suppresses high-order fluctuations and this is needed to produce convex shapes, which requires
\eqS
r^{2}+2r^{\prime2}-rr''\geq0.
\eqE
This condition allows the equation of motion to be conveniently integrated.
\\

The convexity requirement is imposed purely for computational convenience, and the theoretical framework can be trivially extended to non-convex shapes by taking the convex-hull of a general shape. In the region where a flat part is in contact with the plane, we can imagine the contact vector jumps from the two points at the ends of the flat part, leading to an infinite $\dot\theta$. However, $T$ is still well-behaved as the integrand of $T$ is 0 in this part.
\\

In simulations for viscous rolling we use $\sigma=0.08$ and $M=10$. For inertial rolling we use $\sigma=0.12$ and $M=5$.

\subsection{Scaling of the return transition in hysteresis}

In 2D inertial rolling, the cruising-halting transition in the hysteresis loop is a very sharp, but continuous, transition. In the state space $[\theta,\dot{\theta}]$, the cruising trajectory encounters an unstable saddle point and falls into a stable fixed point away from the original trajectory, so the critical behaviour near this transition region is determined by characters of an unstable saddle fixed point. We compute the flow field for each point in the state space $[\theta,\dot{\theta}]$, which is simply $[\dot\theta,\ddot{\theta}]$. Fixed points where the flow vanishes satisfy $\dot\theta=0$, $\ddot\theta=0$, so their locations are simply solutions of (from Equation 9)
\eqS
r(\theta)\sin\alpha-r'(\theta)\cos\alpha=0.
\eqE
We explicitly compute flow fields for $\alpha$ above and below $\alpha^{*}$ and simulate rolling trajectories for a disk with fixed shape, with $\alpha^{*}\approx0.057$ and $m/m_{0}=10$. Initial simulation conditions are $[\pi/2,0.04]$. Results are shown in Figure S2.

\begin{figure}[htb]
\centering
\includegraphics[width=6in]{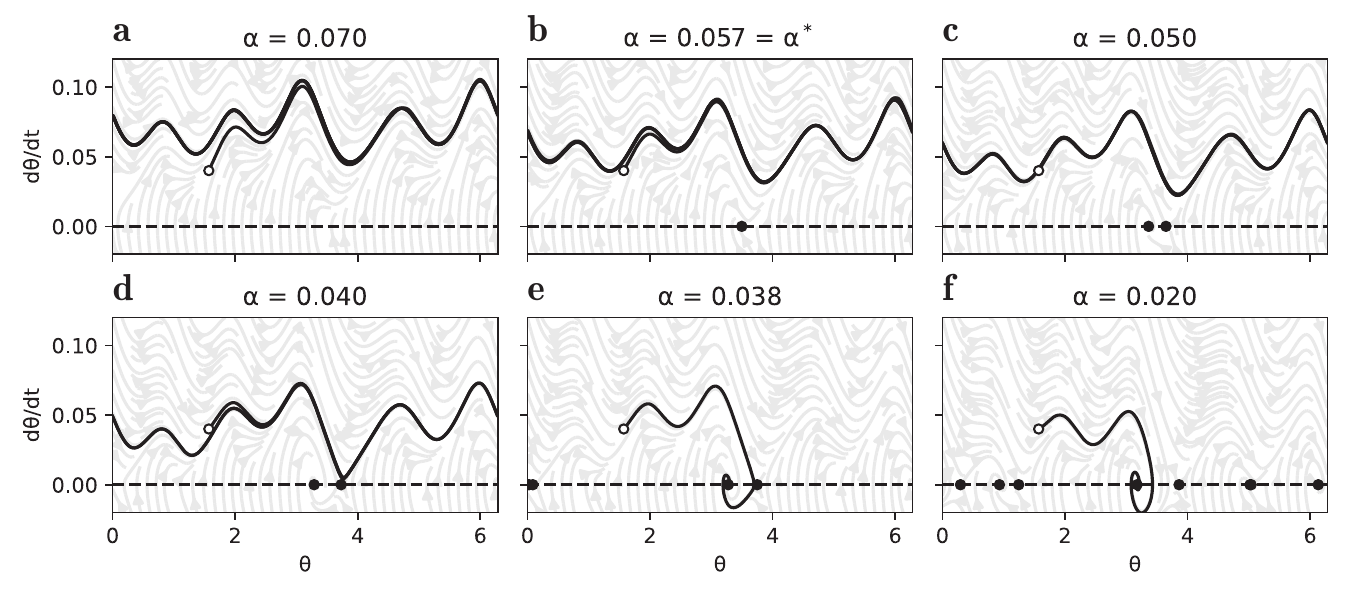}
\caption{Flow field (grey lines) and simulation results (black lines) for inertial rolling of a fixed disk with $\alpha^{*}\approx0.057$, at six different $\alpha$ values. Hollow marker is the initial state, solid markers are fixed points.}
\label{fig:SI_FLOW}
\end{figure}

For $\alpha>\alpha^{*}$ (Figure S2 a), there are no fixed point in the state space so rolling occurs indefinitely. A single fixed point appears at $\alpha=\alpha^{*}$ (Figure S2 b), which splits in two when $\alpha$ decreases slightly (Figure S2 c), but they do not affect the cruising trajectory immediately below $\alpha^{*}$. Instead, near $\alpha=0.039$, a cruising-to-halting transition occurs when the cruising trajectory crosses into the other side of the saddle point (Figure S2 d, e). For small values of $\alpha$, many more fixed points can appear (Figure S2 f). At exactly the transition point, which would correspond to a different critical angle $\alpha^{**}$, we are interested in the dynamics near this saddle point located at $\theta=\theta^{**}$, $\dot\theta=0$. $\dot\theta$ can be written as (for $c_{1}>0$, $c_{2}>0$)
\eqS
\dot\theta(\theta^{**}+\delta\theta)=\begin{cases}-c_{1}\delta\theta&\quad\text{if }\delta\theta\leq0,\\
c_{2}\delta\theta&\quad\text{if }\delta\theta>0.\\
\end{cases}
\eqE
Integrating over $1/\dot\theta$ leads to a divergent integral, so we investigate a case where $\alpha=\alpha^{**}+\delta\alpha$ for some small $\delta\alpha$. We model the crossing behaviour by computing eigenvalues of the matrix
\eqS
\begin{pmatrix}
    -c_{1}\delta\theta&c_{3}\delta\alpha\\
    c_{3}\delta\alpha&c_{2}\delta\theta\\
\end{pmatrix}.
\eqE
The relevant eigenvalue is then used to approximate the behaviour of $\dot\theta$
\eqS
\dot\theta\approx\frac{1}{2}\left[(c_{2}-c_{1})\delta\theta+\sqrt{(c_{1}+c_{2})^{2}\delta\theta^{2}+4c_{3}^{2}\delta\alpha^{2}}\right].
\eqE
We first examine the branch $\delta\theta>0$. We are interested in the integral
\eqS
\int_{0}^{\epsilon}\frac{1}{\frac{1}{2}\left[(c_{2}-c_{1})\delta\theta+\sqrt{(c_{1}+c_{2})^{2}\delta\theta^{2}+4c_{3}^{2}\delta\alpha^{2}}\right]}\dd\delta\theta
\eqE
and when $\delta\alpha=0$, the divergence occurs at small $\delta\theta$ so we expand the denominator in $\delta\theta$ to arrive at
\eqS
\int_{0}^{\epsilon}\frac{1}{c_3\delta\alpha+\frac{1}{2}(c_{2}-c_{1})\delta\theta+\frac{1}{8}\frac{(c_{1}+c_{2})^{2}\delta\theta^{2}}{c_{3}\delta\alpha}+\mathcal{O}(\delta\theta^{3})}\dd\delta\theta.
\eqE
Because $c_{1}\neq c_{2}$ in general, we have encountered a logarithmic integral instead of a quadratic integral. Retaining up to $\delta\theta$ in the denominator and integrating gives
\eqS
\frac{\left[\ln\left(\frac{c_{2}-c_{1}}{2}\delta\theta+c_{3}\delta\alpha\right)\right]_{0}^{\epsilon}}{\frac{c_{2}-c_{1}}{2}}\approx-\frac{2}{c_{2}-c_{1}}\ln\delta\alpha.
\eqE
This means the period $T$ near the return critical angle $\alpha=\alpha^{**}+\delta\alpha$ scales logarithmically with $\delta\alpha$, and as such the rolling speed is
\eqS
\frac{2\pi}{T}\propto-\frac{1}{\ln\delta\alpha}
\eqE
which is an extremely sharp transition, that can appear as discontinuous even in simulations.

\section{3D rolling}

\subsection{Equations of motion}
On a 3D object, torque and force balance, and the non-slip condition, give
\eqS
\begin{split}
I\dot{\bm\omega}&=\bm{r}\cross\bm{f}-\gamma\bm{\omega}\\
m\dot{\bm{v}}&=\bm{f}+\bm{F}\\
\bm{v}&=-\bm\omega\cross\bm{r}.
\end{split}
\eqE 
Combining these into a single equation, we have
\eqS
\begin{split}
I\dot{\bm\omega}=&-\bm{r}\cross(m\dot{\bm\omega}\cross\bm{r}+m\bm\omega\cross\dot{\bm{r}}+\bm{F})-\gamma\bm{\omega}\\
=&-mr^{2}\dot{\bm\omega}+m(\bm{r}\cdot\dot{\bm\omega})\bm{r}-m\bm{r}\cross(\bm\omega\cross\dot{\bm{r}})-\bm{r}\cross\bm{F}-\gamma\bm\omega.
\end{split}
\eqE
It is convenient to first calculate $\bm{r}\cdot\dot{\bm{\omega}}$. This is
\eqS
\begin{split}
I\bm{r}\cdot\dot{\bm{\omega}}=&-\bm{r}\cdot\bm{r}\cross(m\dot{\bm\omega}\cross\bm{r}+m\bm\omega\cross\dot{\bm{r}}+\bm{F})-\gamma\bm{r}\cdot\bm{\omega}\\
\bm{r}\cdot\dot{\bm{\omega}}=&-\frac{\gamma}{I}\bm{r}\cdot\bm{\omega}
\end{split}
\eqE
so now we have an expression for $\dot{\bm\omega}$:
\eqS
\begin{split}
(I+mr^{2})\dot{\bm\omega}&=-\frac{m\gamma}{I}(\bm{r}\cdot\bm\omega)\bm{r}-m\bm{r}\cross(\bm\omega\cross\dot{\bm{r}})-\bm{r}\cross\bm{F}-\gamma\bm\omega\\
\dot{\bm\omega}&=-\frac{1}{I+mr^{2}}(\bm{r}\cross\bm{F}+\gamma\bm\omega)-\frac{1}{I+mr^{2}}\frac{m\gamma}{I}(\bm{r}\cdot\bm\omega)\bm{r}-\frac{m}{I+mr^{2}}\bm{r}\cross(\bm\omega\cross\dot{\bm{r}}).
\end{split}
\eqE
In the viscous regime, setting $m=I=0$ in Equation 34 gives simply
\eqS
\bm\omega=-\frac{1}{\gamma}\bm{r}\cross\bm{F}.
\eqE
\\

\subsection{Quaternion algebra}
Hamilton's quaternion convention $ijk=-1$ is used in this work, and it is worth noting that some literatures use the JPL convention $ijk=1$, which can cause confusion. Detailed discussions of quaternions and these conventions can be found in \cite{Sola2017,Altmann2005}.
\\

A quaternion $q$ that encodes rotational information can be written as a sum of scalar part $q_{s}$ and vector part $\bm{q}$:
\eqS
q=q_{s}+\bm{q}=\cos\frac{\psi}{2}+\sin\frac{\psi}{2}\bm{u}
\eqE
where $\bm{u}$ is a unit vector in lab frame around which the object is rotated by angle $\psi$. A quaternion product between two general quaternions $a=a_{s}+\bm{a}$ and $b=b_{s}+\bm{b}$ is given by
\eqS
ab=(a_{s}b_{s}-\bm{a}\cdot\bm{b})-(a_{s}\bm{b}+b_{s}\bm{a}+\bm{a}\cross\bm{b}).
\eqE
Treating $\bm{r}_{0}$ and $\bm{n}_{0}$, and their lab-frame counterparts as quaternions with zero scalar part, they are related via
\eqS
\begin{split}
\bm{r}=&q\bm{r}_{0}\bar{q}\\
\bm{n}=&q\bm{n}_{0}\bar{q}
\end{split}
\eqE
where quaternion product is assumed. $\bm{r}_{0}$ and $\bm{n}_{0}$ have expressions
\eqS
\begin{split}
\bm{r}_{0}=r(\theta,\phi)\cdot\begin{pmatrix}\sin\theta\cos\phi\\\sin\theta\sin\phi\\\cos\theta\end{pmatrix}\\
\bm{n}_{0}=\frac{\partial_{\theta}\bm{r}_{0}\cross\partial_{\phi}\bm{r}_{0}}{|\partial_{\theta}\bm{r}_{0}\cross\partial_{\phi}\bm{r}_{0}|},
\end{split}
\label{eq:rn}
\eqE
so we can compute $\bm{r}$ and $\bm{n}$ using $q$ directly.
\\

Furthermore, the time evolution of $q$ is closely related to the lab-frame angular velocity $\bm\omega$:
\eqS
\dot{q}=\frac{1}{2}\bm\omega q.
\eqE

\subsection{Contact point time-evolution}

The derivations so far allow us to calculate $\bm\omega$ or its derivative at each time point given $\theta$ and $\phi$. To complete the derivation we need $\dot\theta$ and $\dot\phi$. To this end, we use the final condition that the surface normal $\bm{n}$ at contact in lab frame is always along the $-z$ direction. The equation of motion of $\bm{n}$ is
\eqS
\dot{\bm{n}}=\bm\omega\cross\bm{n}+\begin{pmatrix}\partial_{\theta}\bm{n}&\partial_{\phi}\bm{n}\end{pmatrix}\begin{pmatrix}\dot{\theta}\\\dot{\phi}\end{pmatrix}
\eqE
and values for $\dot\theta$, $\dot\phi$ can be obtained by setting the above to 0, and making sure $\bm{n}=(0,0,-1)^{T}$ at initialisation. The normal contact condition is thus
\eqS
-\begin{pmatrix}\partial_{\theta}\bm{n}&\partial_{\phi}\bm{n}\end{pmatrix}\begin{pmatrix}\dot{\theta}\\\dot{\phi}\end{pmatrix}=\bm\omega\cross\bm{n}
\eqE
and dot-multiplying both sides by $\partial_{\theta}\bm{r}$ and $\partial_{\phi}\bm{r}$:
\eqS
\begin{pmatrix}
L_{2}&M_{2}\\
M_{2}&N_{2}
\end{pmatrix}\begin{pmatrix}\dot{\theta}\\\dot{\phi}\end{pmatrix}=\begin{bmatrix}\partial_{\theta}\bm{r}\cdot(\bm\omega\cross\bm n)\\\partial_{\phi}\bm{r}\cdot(\bm\omega\cross\bm n)\end{bmatrix}
\eqE
where $L_{2}$, $M_{2}$, and $N_{2}$ are coefficients of the second fundamental form
\eqS
\begin{split}
L_{1}=&-\partial_{\theta}\bm{r}\cdot\partial_{\theta}\bm{n}=-\partial_{\theta}\bm{r}_{0}\cdot\partial_{\theta}\bm{n}_{0}\\
M_{1}=&-\partial_{\theta}\bm{r}\cdot\partial_{\phi}\bm{n}=-\partial_{\theta}\bm{r}_{0}\cdot\partial_{\phi}\bm{n}_{0}\\
N_{1}=&-\partial_{\phi}\bm{r}\cdot\partial_{\phi}\bm{n}=-\partial_{\phi}\bm{r}_{0}\cdot\partial_{\phi}\bm{n}_{0}.\\
\end{split}
\eqE
To proceed, we write $\bm{n}$ as
\eqS
\bm{n}=\frac{\partial_{\theta}\bm{r}\cross\partial_{\phi}\bm{r}}{\sqrt{E_{1}G_{1}-F_{1}^{2}}}
\eqE
where $E_{1}$, $F_{1}$, and $G_{1}$ are coefficients of the first fundamental form
\eqS
\begin{split}
E_{1}=&\partial_{\theta}\bm{r}\cdot\partial_{\theta}\bm{r}=\partial_{\theta}\bm{r}_{0}\cdot\partial_{\theta}\bm{r}_{0}\\
F_{1}=&\partial_{\theta}\bm{r}\cdot\partial_{\phi}\bm{r}=\partial_{\theta}\bm{r}_{0}\cdot\partial_{\phi}\bm{r}_{0}\\
G_{1}=&\partial_{\phi}\bm{r}\cdot\partial_{\phi}\bm{r}=\partial_{\phi}\bm{r}_{0}\cdot\partial_{\phi}\bm{r}_{0}\\
\end{split}
\eqE
and using
\eqS
\begin{split}
\bm{n}\cross\partial_{\theta}\bm{r}=&\frac{E_{1}\partial_{\phi}\bm{r}-F_{1}\partial_{\theta}\bm{r}}{\sqrt{E_{1}G_{1}-F_{1}^{2}}}\\
\bm{n}\cross\partial_{\phi}\bm{r}=&\frac{F_{1}\partial_{\phi}\bm{r}-G_{1}\partial_{\theta}\bm{r}}{\sqrt{E_{1}G_{1}-F_{1}^{2}}}
\end{split}
\eqE
we get
\eqS
\begin{split}
\begin{pmatrix}
L_{2}&M_{2}\\
M_{2}&N_{2}
\end{pmatrix}\begin{pmatrix}\dot{\theta}\\\dot{\phi}\end{pmatrix}=\frac{1}{\sqrt{E_{1}G_{1}-F_{1}^{2}}}\begin{pmatrix}-F_{1} & E_{1}\\-G_{1} & F_{1}\end{pmatrix}\begin{pmatrix}\bm\omega\cdot\partial_{\theta}\bm{r}\\\bm\omega\cdot\partial_{\phi}\bm{r}\end{pmatrix}.
\end{split}
\eqE
Numerical integration can now be formed to propagate $\theta$ and $\phi$.
\\

It is convenient to first calculate the coefficients of the fundamental forms in terms of $r(\theta,\phi)$ using Eq.~\eqref{eq:rn}. Here we use the short hand $r_{\theta}\equiv\frac{\partial}{\partial\theta}r(\theta,\phi)$ and $r_{\phi}\equiv\frac{\partial}{\partial\phi}r(\theta,\phi)$, giving
\eqS
\begin{split}
E_{1}=&r^2+r_{\theta }^2\\
F_{1}=&r_{\theta } r_{\phi }\\
G_{1}=&r^2 \sin ^2\theta+r_{\phi }^2\\
L_{2}=&\frac{-r \sin \theta \left[2 r_{\theta }^2+r \left(r-r_{\theta \theta }\right)\right]}{\sqrt{E_{1}G_{1}-F_{1}^{2}}}\\
M_{2}=&\frac{r \left[r r_{\theta \phi } \sin \theta-r_{\phi } \left(2 r_{\theta } \sin \theta+r \cos \theta\right)\right]}{\sqrt{E_{1}G_{1}-F_{1}^{2}}}\\
N_{2}=&\frac{-r \sin \theta \left(-r r_{\theta } \sin \theta \cos \theta+r \left(r \sin ^2\theta-r_{\phi \phi }\right)+2 r_{\phi }^2\right)}{\sqrt{E_{1}G_{1}-F_{1}^{2}}}
\end{split}
\eqE

\subsection{Viscous rolling quality factor}
\label{3DHalt}
In the viscous limit, $\bm\omega$ is 
\eqS
\bm\omega=-\frac{1}{\gamma}\bm{r}\cross\bm{F}
\eqE
and a halting condition can be worked out just as in the 2D case. For an equilibrium orientation of the sphere to exist (stable or unstable), the displacement vector must be parallel to $\bm{F}$ at some point. A stable equilibrium state cannot exist if
\eqS
\frac{\bm{r}}{r}\cdot\bm{n}\leq\cos\alpha
\eqE
which is
\eqS
\frac{r}{\sqrt{r^{2}+\frac{(\partial_{\phi}r)^{2}}{\sin^{2}\theta}+(\partial_{\theta}r)^{2}}}\leq\cos\alpha.
\eqE
To re-cast it to the form similar to the 2D case, we wish to obtain an expression relating $r$ to $\tan\alpha$. Simple manipulations give
\eqS
\frac{|\bm{\nabla}_{s}r|}{r}\leq\tan\alpha
\eqE
or to write it out in full:
\eqS
Q\equiv\frac{|\bm{\nabla}_{s}r|}{r}=\frac{\sqrt{r_{\theta}^{2}+r_{\phi}^{2}/\sin^{2}\theta}}{r}=|\bm{\nabla}_{s}\ln r|
\eqE
where $\bm{\nabla}_{s}$ is the spherical part of the gradient operator, and $\bar{q}$ is
\eqS
\bar{q}=\max_{\theta,\phi}Q(\theta,\phi).
\eqE

\subsection{3D halting states and stability analysis}
\label{3DHaltAnalysis}
Halting states are contact points $(\theta_{H},\phi_{H})$ that satisfy, for a given $\alpha$, 
\eqS
Q(\theta_{H},\phi_{H})=\tan\alpha,
\eqE
and for each point $(\theta_{H},\phi_{H})$ we need to additionally specify how the sphere is to be rotated on the spot to align $\bm{r}$ to $\bm{F}$, but this degree of freedom is fixed once $(\theta_{H},\phi_{H})$ is fixed. As such, for each point $(\theta_{H},\phi_{H})$, there corresponds to a point in $SO(3)$ where the external torque is zero. The connected set of $(\theta_{H},\phi_{H})$ is a 1D loop, and this is easy to see since the condition $Q(\theta_{H},\phi_{H})=\tan\alpha$ constructs level sets in the 2D $\theta-\phi$  space which are in general 1D loops. This 1D loop in the $\theta-\phi$ space acquires the additional on-the-spot rotation angle to generate a 1D loop in $SO(3)$, so the set of halting states are 1D loops. 
\\

The stability of halting states is also of interest. We limit discussions to the viscous regime, so that the angular velocity is
\eqS
\bm\omega=-\frac{1}{\gamma}\bm{r}\cross\bm{F}
\eqE
and in a halting state, $\bm{r}\cross\bm{F}=\bm{0}$. We apply a small perturbation $\delta q$ to the orientation $q$, we can define the small perturbation vector $\bm\mu$ as
\eqS
\delta q\bar{q}\equiv\frac{1}{2}\bm\mu
\eqE
such that the perturbed equation of motion for $q$ reads
\eqS
\delta \dot{q}=\frac{1}{2}\delta\bm\omega q+\frac{1}{2}\bm\omega\delta q
\eqE
Re-arranging, we have
\eqS
\begin{split}
\delta\dot{q}\bar{q}&=\frac{1}{2}\delta\bm\omega+\frac{1}{2}\bm\omega\delta q\bar{q}\\
\frac{\dd}{\dd t}(\delta q\bar{q})-\delta q\dot{q}^{*}&=\frac{1}{2}\delta\bm\omega+\frac{1}{4}\bm\omega\bm\mu\\
\frac{1}{2}\dot{\bm{\mu}}+\frac{1}{4}\bm\mu\bm\omega&=\frac{1}{2}\delta\bm\omega+\frac{1}{4}\bm\omega\bm\mu\\
\dot{\bm{\mu}}&=\delta\bm\omega+\bm\omega\cross\bm\mu
\end{split}
\eqE
and we wish to know whether the magnitude of $\bm\mu$ grows or shrinks over time. Doing a dot product:
\eqS
\frac{\dd}{\dd t}|\bm{\mu}|^{2}=2\delta\bm\omega\cdot\bm\mu=-\frac{2}{\gamma}(\delta\bm{r}\cross\bm{F})\cdot\bm\mu
\eqE
which is quadratic in $\bm\mu$. To see this, notice $\bm{r}$, $\theta$, and $\phi$ change by
\eqS
\begin{split}
\delta\bm{r}=&\bm\mu\cross\bm{r}+\begin{pmatrix}\partial_\theta\bm{r}&\partial_\phi\bm{r}\end{pmatrix}\begin{pmatrix}\delta \theta\\\delta \phi\end{pmatrix}\\
\begin{pmatrix}\delta \theta\\\delta \phi\end{pmatrix}=&\frac{\begin{pmatrix}
L_{2}&M_{2}\\
M_{2}&N_{2}
\end{pmatrix}^{-1}\begin{pmatrix}-F_{1} & E_{1}\\-G_{1} & F_{1}\end{pmatrix}\begin{pmatrix}\bm\mu\cdot\partial_{\theta}\bm{r}\\\bm\mu\cdot\partial_{\phi}\bm{r}\end{pmatrix}}{\sqrt{E_{1}G_{1}-F_{1}^{2}}}
\end{split}
\eqE
so $\delta\bm{r}$ is in fact linear in $\bm\mu$ and $\frac{\dd}{\dd t}|\bm\mu|^{2}$ is a quadratic. Stability of the state can then be calculated in general by diagonalising the matrix.
\\

A simplifying scenario is where $\bm\mu=\mu\bm{n}$ for some small value $\mu$. This is a perturbation where the sphere is spun on the spot, and from the definition of $\bm{n}$ we have $\bm\mu\cdot\partial_{\theta}\bm{r}=0$ so $\delta\theta=\delta\phi=0$. We then have
\eqS
\frac{\gamma}{2}\frac{\dd}{\dd t}|\bm{\mu}|^{2}=-(\delta\bm{r}\cross\bm{F})\cdot\bm\mu=-\mu^{2}[(\bm{n}\cross\bm{r})\cross\bm{F}]\cdot\bm{n}
\eqE
and in a halting state we know $\bm{F}=\frac{F}{r}\bm{r}$ so, up to a positive multiplicative constant,
\eqS
\begin{split}
\frac{\gamma}{2}\frac{\dd}{\dd t}|\bm{\mu}|^{2}\propto&-[(\bm{n}\cross\bm{r})\cross\bm{r}]\cdot\bm{n}\\
=&[r^{2}\bm{n}-(\bm{r}\cdot\bm{n})\bm{r}]\cdot\bm{n}\\
=&r^{2}-(\bm{r}\cdot\bm{n})^2\\
=&r^{2}-r^{2}\cos^{2}\alpha\geq0
\end{split}
\eqE
so spinning the sphere on the spot is always unstable except if $\alpha=0$.
\\

Having studied the case $\bm{\mu}$ parallel to $\bm{n}$, we now look at the orthogonal directions $\partial_{\theta}\bm{r}$ and $\partial_{\phi}\bm{r}$. Without loss of generality, we set $\bm\mu=\mu\partial_{\theta}\bm{r}$ so
\eqS
\begin{split}
\frac{\gamma}{2}\frac{\dd}{\dd t}|\bm{\mu}|^{2}=&-\frac{\mu F}{r}(\delta\bm{r}\cross\bm{r})\cdot\partial_{\theta}\bm{r}\\
=&-\frac{\mu F}{r}[(\mu\partial_{\theta}\bm{r}\cross\bm{r}+\delta\phi\cdot\partial_{\phi}\bm{r})\cross\bm{r}]\cdot\partial_{\theta}\bm{r}\\
=&\frac{\mu^2 F}{r}\left[E_{1}r^{2}-(\bm{r}\cdot\partial_{\theta}\bm{r})^{2}-\frac{\delta\phi}{\mu}(\partial_{\phi}\bm{r}\cross\bm{r})\cdot\partial_{\theta}\bm{r}\right]\\
=&\frac{\mu^2 F}{r}\left[E_{1}r^{2}-(\bm{r}\cdot\partial_{\theta}\bm{r})^{2}-\frac{\delta\phi}{\mu}(\partial_{\theta}\bm{r}\cross\partial_{\phi}\bm{r})\cdot\bm{r}\right]
\end{split}
\eqE
\\

The contact angle changes are
\eqS
\begin{split}
\begin{pmatrix}\delta \theta\\\delta \phi\end{pmatrix}=&\frac{\begin{pmatrix}
L_{2}&M_{2}\\
M_{2}&N_{2}
\end{pmatrix}^{-1}\begin{pmatrix}-F_{1} & E_{1}\\-G_{1} & F_{1}\end{pmatrix}\begin{pmatrix}\mu E_{1}\\\mu F_{1}\end{pmatrix}}{\sqrt{E_{1}G_{1}-F_{1}^{2}}}\\
=&\frac{\mu}{K\sqrt{E_{1}G_{1}-F_{1}^{2}}}\begin{pmatrix}M_{2}\\-L_{2}\end{pmatrix}
\end{split}
\eqE
where $K\equiv\frac{L_{2}N_{2}-M_{2}^{2}}{E_{1}G_{1}-F_{1}^{2}}$ is the Gaussian curvature. Direct substitution leads to
\eqS
\begin{split}
\frac{\gamma}{2}\frac{\dd}{\dd t}|\bm{\mu}|^{2}=&\frac{\mu^2 F}{r}\left[E_{1}r^{2}-(\bm{r}\cdot\partial_{\theta}\bm{r})^{2}+\frac{L_{2}}{K}\bm{n}\cdot\bm{r}\right]\\
=&\frac{\mu^2 F}{r}\left[E_{1}r^{2}-(\bm{r}\cdot\partial_{\theta}\bm{r})^{2}+\frac{rL_{2}}{K}\cos\alpha\right].
\end{split}
\eqE
Notice that the sum of the first two terms is always positive, serving as a destabilising effect. For the second term, $L_{2}<0$ in general so it is a stabilising term, where the contact point moves to prevent toppling. 

\subsection{Simulation}
\label{3DSIM}
To generate a random sphere, we write the radius in the body frame as the real part of a sum of spherical harmonics
\eqS
r(\theta,\phi)=1+\text{Re}\left[\sum_{l=2}^{5}\sum_{m=-l}^{l}\frac{A^{m}_{l}}{l(l+1)}Y^{m}_{l}(\theta,\phi)\right]
\eqE
where again $A_{l}^{m}\sim\mathcal{N}(0,\sigma)$, with $\sigma=0.25$, and $Y_{l}^{m}$ are standard spherical harmonic functions. We use Scipy's initial-value-problem integrator with explicit Runge-Kutta method of order 8 to integrate the equations of motion for $q$ and $\theta$, $\phi$. Numerical errors are checked by outputting the normalisation of the unit quaternion $q$ as well as the orientation of the lab frame normal vector $\bm{n}$.
\\

To generate different initial sphere orientations, we first pick a random contact point $(\theta,\phi)$ and generate a quaternion that puts that point onto contact with the rolling plane along the shortest path. We then spin the sphere on the spot by a random angle, represented by a second quaternion. Concatenating these two quaternions gives the initial quaternion value for further computation.

\subsection{Two stable orbits in viscous rolling}
\label{3DTube}
This sub-section provides a visual explanation of shapes of the 2 stable viscous rolling orbits in $SO(3)$. Conceptually, we can imagine the process of creating surface irregularities as continuously changing the radii values of random patches on an initially isotropic sphere, and by computing trajectories passing through a particular point on the 3D shape we can visualise the set of orientation states affected by a single point perturbation. 
\\

\begin{figure}[htb]
\centering
\includegraphics[width=3in]{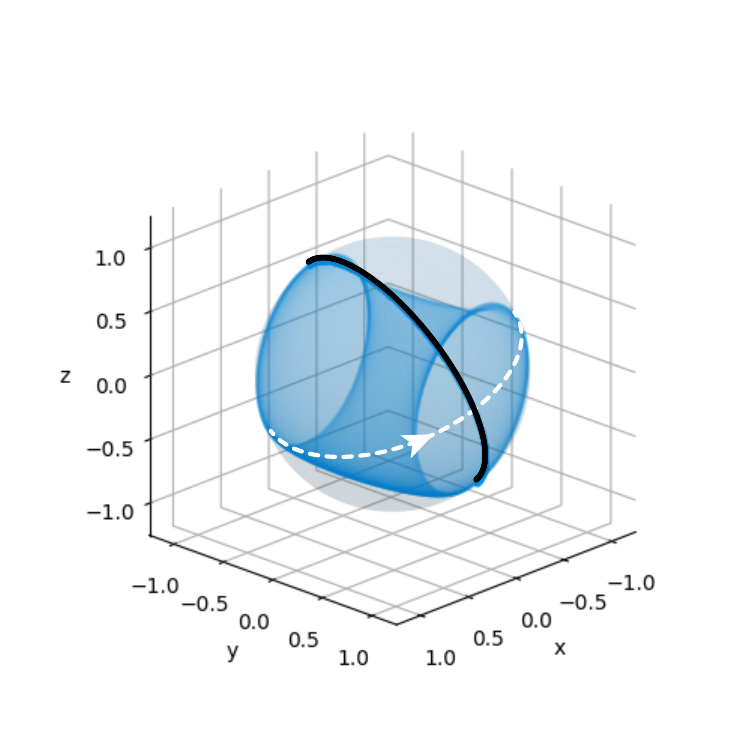}
\caption{Visualising surface perturbation in $SO(3)$. A perfectly round sphere is generated and the black solid line represents the set of orientations that share a common, arbitrarily chosen contact point, and ends of the line are antipodes of each other, forming a closed loop. Simulations are then performed with the sphere initially in contact with the plane on this point, but spun on the spot by angles between 0 and 360$\degree$. The white dashed line represents a single simulation trajectory for illustrative purposes, and when all trajectories are plotted in light blue colour a tube is formed.}
\label{fig:SI_barrel}
\end{figure}

To this end, we generate a perfect sphere and pick an arbitrary point on the sphere's surface, and orient the perfect sphere such that the picked point is in contact with the plane. We then generate a series of orientations by spinning the sphere on the spot, such that the same point is always in contact but the sphere would roll over it from different directions. The set of initial orientations can be represented as a closed segment in $SO(3)$ since the original orientation is recovered after a 360$\degree$ spin. Performing the simulations with this series of initial orientations allows us to obtain quaternion representations of all orientations from trajectories passing though the picked point, and when plotted together they mimic a tube running though the centre of the $SO(3)$ 3-ball (Fig \ref{fig:SI_barrel}). This tube is a closed surface, since all trajectories are closed when a perfect sphere is rolled, and 2 sub-spaces are thus formed. One of these spaces goes through the middle of the sphere while the other wraps around the tube on the outside, and the 2 closed orbits that eventually formed have shapes originating from these sub-spaces.

\subsection{Rolling regimes of 3D experiments}
\label{3DExptMass}

In experiments, we can deduce the mass unit simply using the rolling speed. The angle $\alpha$ is approximately $\sin\alpha\approx0.20$, and $F=mg$ with $g$ the gravitational acceleration. In steady state, we expect the input energy from gravity to balance the dissipation from the viscosity. Denoting the speed in the $x$-direction as $v_{x}$, the rate of energy input is
\eqS
\dot{E}_{g}=mgv_{x}\sin\alpha.
\eqE
On the other hand, the energy dissipation from $\gamma$ is 
\eqS
\dot{E}_{\gamma}=\gamma\omega^{2}
\eqE
where we estimate the angular velocity as
\eqS
\omega=v_{x}/\expval{r}
\eqE
so that equating the two powers gives
\eqS
\gamma v_{x}^{2}=mgv_{x}\expval{r}^{2}\sin\alpha.
\eqE
This leads to an expression for the ratio
\eqS
\frac{m}{\gamma}=\frac{v_{x}}{g\expval{r}^{2}\sin\alpha}.
\eqE
\\

The dimensionless mass scale is given by the quantity
\eqS
\frac{m}{\frac{\gamma^{2}}{F\expval{r}^{3}}}=\frac{m^{2}g\expval{r}^{3}}{\gamma^{2}}
\eqE
and substituting in the expression for $m/\gamma$ gives
\eqS
\frac{m}{\frac{\gamma^{2}}{F\expval{r}^{3}}}=g\expval{r}^{3}\cdot\frac{v_{x}^{2}}{g^{2}\expval{r}^{4}\sin^{2}\alpha}=\frac{v_{x}^{2}}{g\expval{r}\sin^{2}\alpha}.
\eqE
\\

In the viscous rolling experiment, we have $\expval{r}\approx10^{-2}$ m, $g=10$ ms$^{-2}$, $v_{x}\approx10^{-2}$ ms$^{-1}$, and $\sin\alpha\approx0.2$. This leads to a reduced mass of $0.025$ so the system is well within the viscous regime.
\\

In the inertia rolling experiment, the speed is instead $v_{x}\approx10^{-1}$ ms$^{-1}$, leading to reduced mass of $2.5$, so inertia effect starts kicking in.
\\













\bibliography{library}